\documentclass[aps,prx,twocolumn,superscriptaddress,nofootinbib]{revtex4-2}
\usepackage{mathrsfs}
\usepackage{amsmath,amssymb,bm,mathtools}
\usepackage{blindtext}
\usepackage{graphicx}
\graphicspath{ {figures/}{figures/eps/}{figures/pdf/} }
\usepackage{array}
\usepackage{float}
\usepackage[export]{adjustbox}
\usepackage{framed}
\usepackage{physics}
\usepackage{xcolor}
\usepackage[justification=raggedright, font=small]{caption}
\usepackage{makecell}
\setcellgapes{4pt}   % vertical padding inside cells
\makegapedcells
\usepackage[caption=false]{subfig}
\usepackage{dcolumn}
\usepackage{bm}% bold math
\usepackage{textcomp}
\usepackage{braket}
\usepackage{comment}
\usepackage{siunitx}
\usepackage{upgreek}
\usepackage{mathrsfs}
\usepackage{multirow}
\usepackage{natbib,hyperref}

\usepackage{booktabs}
\usepackage{wasysym}
\usepackage{comment}
\usepackage{times}

\def\pso{$\text{Pr}_2\text{Sn}_2\text{O}_7$}
\def\pzo{$\text{Pr}_2\text{Zr}_2\text{O}_7$}
\def\pho{$\text{Pr}_2\text{Hf}_2\text{O}_7$}
\def\hto{$\text{Ho}_2\text{Ti}_2\text{O}_7$}
\def\dto{$\text{Dy}_2\text{Ti}_2\text{O}_7$}

\def\yto{$\text{Yb}_2\text{Ti}_2\text{O}_7$}

\def\czo{$\text{Ce}_2\text{Zr}_2\text{O}_7$}
\def\cso{$\text{Ce}_2\text{Sn}_2\text{O}_7$}
\def\cho{$\text{Ce}_2\text{Hf}_2\text{O}_7$}

\begin{document}

\title{Disorder-induced proximate quantum spin ice phase in \texorpdfstring{\pso}{Pr2Sn2O7}}

\author{Yi~Luo}
\email{luoy1@ornl.gov}
\email{luoyi1991.good@gmail.com}
\affiliation{Neutron Scattering Division, Oak Ridge National Laboratory, Oak Ridge, Tennessee 37831, USA}
\author{Brenden~R.~Ortiz}
%\email{ortizbr@ornl.gov}
\affiliation{Materials Science and Technology Division, Oak Ridge National Laboratory, Oak Ridge, Tennessee 37831, USA}
\author{Miles~J.~Knudtson}
\affiliation{Materials Department and California Nanosystems Institute, University of California Santa Barbara, Santa Barbara, California 93106, USA}
\author{Stephen~D.~Wilson}
\affiliation{Materials Department and California Nanosystems Institute, University of California Santa Barbara, Santa Barbara, California 93106, USA}
\author{Jue~Liu}
\affiliation{Neutron Scattering Division, Oak Ridge National Laboratory, Oak Ridge, Tennessee 37831, USA}
\author{Benjamin~A.~Frandsen}
\affiliation{Department of Physics and Astronomy, Brigham Young University, Provo, Utah 84602, USA}
\author{Si~Athena~Chen}
\affiliation{Neutron Scattering Division, Oak Ridge National Laboratory, Oak Ridge, Tennessee 37831, USA}
\author{Matthias~D.~Frontzek}
\affiliation{Neutron Scattering Division, Oak Ridge National Laboratory, Oak Ridge, Tennessee 37831, USA}
\author{Andrey~A.~Podlesnyak}
\affiliation{Neutron Scattering Division, Oak Ridge National Laboratory, Oak Ridge, Tennessee 37831, USA}
\author{Joseph~A.~M.~Paddison}
\email{paddisonja@ornl.gov}
\affiliation{Neutron Scattering Division, Oak Ridge National Laboratory, Oak Ridge, Tennessee 37831, USA}
\author{Adam~A.~Aczel}
\email{aczelaa@ornl.gov}
\affiliation{Neutron Scattering Division, Oak Ridge National Laboratory, Oak Ridge, Tennessee 37831, USA}
\date{\today}

\begin{abstract}
Magnetic pyrochlores with non-Kramers rare-earth ions provide a platform for exploring emergent gauge physics and quantum spin-ice behavior, yet the influence of structural disorder on their ground states remains insufficiently understood. Here we combine bulk characterization and single-crystal neutron-scattering measurements to investigate the non-Kramers pyrochlore Pr$_2$Sn$_2$O$_7$. At temperatures below $\sim1$~K, the system exhibits key hallmarks of quantum spin-ice physics, including anisotropic spin-ice correlations and two distinct dynamical timescales. Upon further cooling, however, we observe a complete spin-freezing transition at $T_f \approx 0.15$~K, accompanied by recovery of the full nuclear Schottky anomaly, the emergence of a gapped magnetic excitation, and the development of incipient $(100)$ magnetic correlations. Comparison with related Pr-based pyrochlores places \pso\ near the spin-frozen boundary of a disorder-perturbed phase diagram. These results establish a disorder-driven framework for the evolution of quantum spin-ice behavior into frozen ground states, revealing how signatures of a proximate quantum spin liquid can persist despite disorder-induced spin freezing in non-Kramers pyrochlores.
\end{abstract}

\maketitle 
\section{Introduction}
The discovery of classical spin ice (CSI) phases in rare-earth pyrochlores~\cite{DenHertog2000,Fennell2009} represents a key milestone in frustrated magnetism, with potential implications for quantum information and magnetic storage~\cite{Broholm2020,Yao2013}. In these systems, strong Ising anisotropy and effective ferromagnetic nearest-neighbor interactions stabilize a ``two-in, two-out'' ground state, with magnetic moments aligned along local $\langle111\rangle$ directions. This configuration gives rise to an extensively degenerate manifold~\cite{ramirez1999zero} and dipolar spin correlations that manifest as pinch-point features in neutron scattering~\cite{Fennell2009}. The inclusion of transverse exchange interactions introduces quantum fluctuations, leading to quantum spin ice (QSI), in which tunneling between ice-rule configurations stabilizes a $U(1)$ quantum spin liquid (QSL)~\cite{hermele2004pyrochlore,Savary2012}. Beyond spinon excitations inherited from CSI, the QSI phase supports emergent gauge excitations, including gapless photon-like modes and gapped topological defects (visons, or ``magnetic monopoles'')~\cite{gingras2014quantum,Chen2016MonopoleCondensation,Chen2017DiracMonopoles}. Early discussions of QSI focused on materials such as \yto~\cite{Ross2011,Scheie2020,robert2015spin}. More recently, Ce-based pyrochlores, such as \czo~\cite{Gaudet2019_Ce2Zr2O7_QSI_PRL,bhardwaj2022sleuthing,Gao2019Ce2Zr2O7,Smith2022U1pi,Smith2025_Ce2Zr2O7_PRX,gao2025neutron}, \cso~\cite{Sibille2019_Ce2Sn2O7_octupole_NatPhys,Poree2025FractionalCe2Sn2O7,Yahne2024_Ce2Sn2O7_PRX}, and \cho~\cite{Poree2023Ce2Hf2O7,smith2025two}, have emerged as particularly promising platforms for realizing quantum spin-liquid physics.\par

\begin{figure}[t] 
\centering
\includegraphics{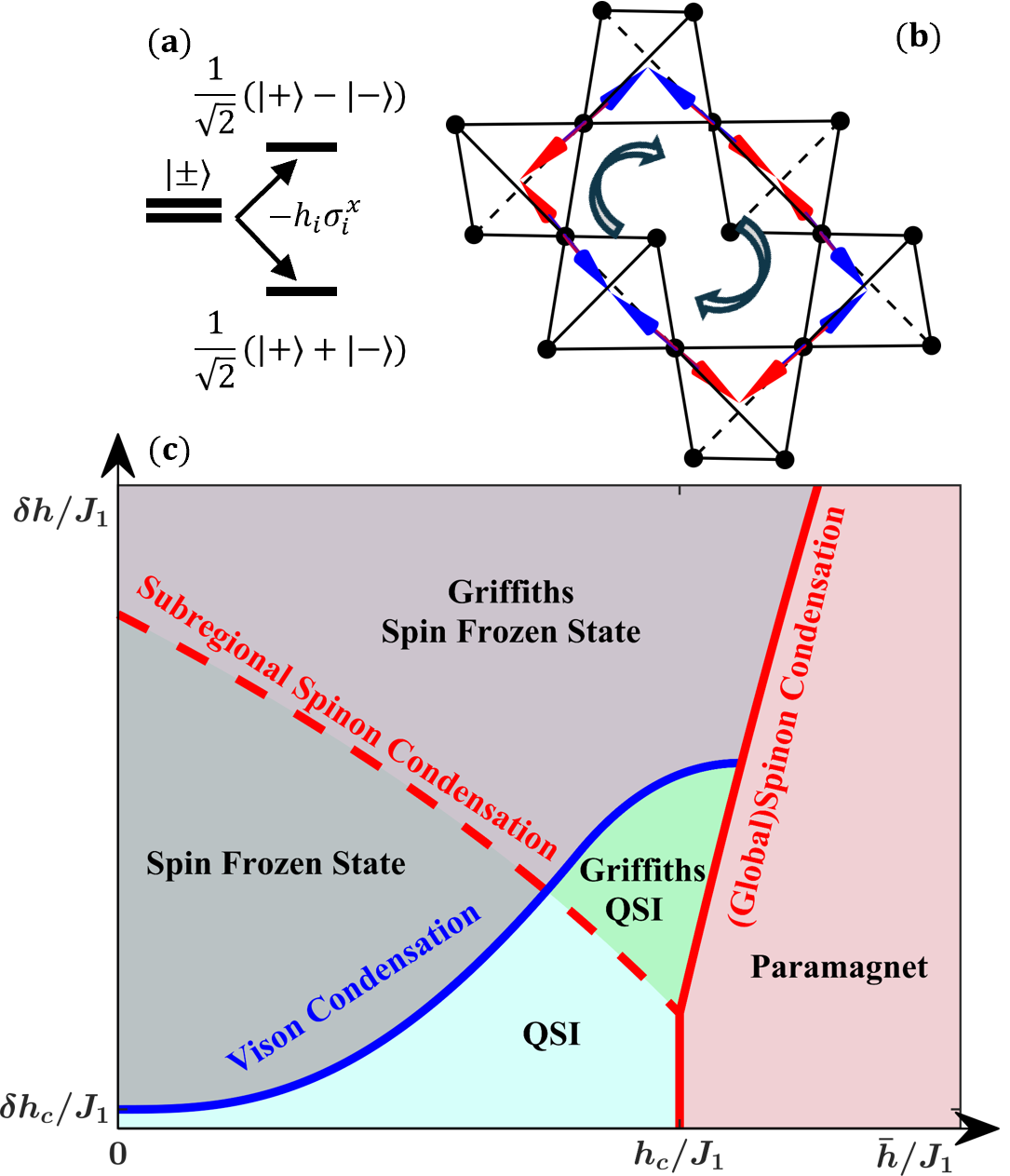}
\caption{\textbf{Conceptual framework for disorder-perturbed non-Kramers quantum spin ice.} 
(a,b) Schematics of (a) the splitting of a non-Kramers doublet by a site-dependent transverse field $h_i$, and (b) a ring-exchange process that flips spins (blue to red) around a hexagonal plaquette. 
(c) Conceptual phase diagram of the antiferromagnetic Ising model with random transverse fields [Eq.~\ref{eq1}], using the nearest-neighbor interaction $J_1$ as the reference energy scale, with weak transverse interactions $|J_{\pm}| \ll J_1$ treated perturbatively, while the symmetry-allowed $J_{\pm\pm}$ and further neighbor terms $J_n$ ($n>1$) are neglected. The parameter $\bar{h}$ denotes the spatial average of $h_i$, while $\delta h$ characterizes the width of its distribution.}
\label{fig:Fig1}
\vspace{-10pt}
\end{figure}

Pr-based pyrochlores constitute another family of promising QSI candidates, featuring a non-Kramers ground-state doublet well separated from the first excited crystal-field (CF) level ($\gtrsim 10$~meV)~\cite{Kimura2013,Anand2016,Princep2013Pr2Sn2O7_CrystalField}. Their low-temperature properties are thus governed by an effective pseudospin-$1/2$ degree of freedom on each Pr$^{3+}$ site. The symmetry-allowed Hamiltonian in the local coordinate frame takes the form
\begin{equation}
\begin{split}
\mathcal{H} &= \sum_{\langle ij \rangle, n} J_n \sigma_i^z \sigma_j^z + \sum_{\langle ij \rangle, 1} \left\{ -J_{\pm}(\sigma^+_i\sigma^-_j + \sigma^-_i\sigma^+_j) \right. \\
&\left. + J_{\pm\pm}\left[\gamma_{ij}\sigma^+_i\sigma^+_j + \gamma_{ij}^*\sigma^-_i\sigma^-_j\right] \right\} - \sum_i h_i \sigma_i^x,
\end{split}
\label{eq1}
\end{equation}
Here, $\sigma_i^{x,y,z}$ are Pauli matrices. The component $\sigma_i^z$ denotes the longitudinal dipole moment along $\langle111\rangle$, and nearest-neighbor Ising interactions $J_1>0$ stabilize the ``two-in, two-out'' state, while further-neighbor longitudinal couplings ($n>1$) are included for completeness. The transverse components $\sigma_i^{x,y}$ (with $\sigma_i^{\pm}\equiv\sigma_i^x\pm i\sigma_i^y$) correspond to quadrupolar degrees of freedom~\cite{Princep2013Pr2Sn2O7_CrystalField,Petit2016}, while symmetry-allowed nearest-neighbor transverse interactions $J_{\pm}$ and $J_{\pm\pm}$ introduce quantum fluctuations~\cite{Ross2011,Petit2016}. In real Pr-based systems, the non-Kramers ground-state doublet can be split by perturbations or disorder that break the local $D_{3d}$ symmetry. This effect is captured by a site-dependent \textit{transverse field} $h_i$ coupled to the quadrupolar component $\sigma_i^x$~\cite{Wen2017,Martin2017,Benton2018,dun2020quantum}[Fig.~\ref{fig:Fig1}(a)].\par

In the perturbative limit of the nearest-neighbor model, where $|J_{\pm}|, |h_i| \ll J_1$ and $J_{\pm\pm}$ is neglected as a subleading term, the low-energy dynamics is governed by an effective ring-exchange process [Fig.~\ref{fig:Fig1}(b)],
\begin{equation}
\mathcal{H}_{\text{ring}} = -\sum_{\hexagon}
\left(K^{J}+K^{h}_{ijklmn}\right)\sigma^{+}_i \sigma^{-}_j \sigma^{+}_k \sigma^{-}_l \sigma^{+}_m \sigma^{-}_n + \text{H.c.},
\label{Hring}
\end{equation}
stabilizing a non-Kramers QSI phase. The uniform contribution $K^{J} = 12 J_{\pm}^3 / J_1^2$ arises from transverse exchange~\cite{hermele2004pyrochlore,Benton2012}, while the plaquette-dependent term $K^{h}_{ijklmn} = 63\, h_i h_j h_k h_l h_m h_n / 16 J_1^5$ originates from site-dependent transverse fields~\cite{Savary2017,Benton2018}.

In the non-Kramers QSI regime, the neutron-scattering response is theoretically predicted to be dominated by photon-like excitations, with a vanishing spinon contribution~\cite{Chen2017DiracMonopoles,an2025quantum}, in contrast to experimental observations in Pr-based QSI candidates. In \pzo, more than 90\% of the spectral weight resides in a broad inelastic continuum~\cite{Kimura2013,Petit2016}, later attributed to disorder-induced transverse fields and transitions within split ground-state doublets~\cite{Wen2017,Martin2017}. By comparison, \pho\ displays enhanced quasielastic spectral weight and a reduced inelastic response forming a gapped excitation near $\hbar\omega \approx 0.2$~meV~\cite{Sibille2016,Sibille2018}. In both compounds, the inelastic magnetic scattering displays a characteristic ``star-fish-like''~\cite{Wen2017} $\boldsymbol{Q}$ dependence devoid of pinch-point features. Furthermore, pronounced sample dependence observed in low-temperature heat capacity and susceptibility for \pzo~\cite{Kimura2013,Tang2022_spinorbital}, \pho~\cite{Sibille2016,Anand2016}, and \pso~\cite{Matsuhira2004,Ortiz2024Pr2Sn2O7}, points to a dominant role of disorder over intrinsic transverse interactions.\par

Although a complete theoretical framework for disordered QSIs remains lacking, the experimental observations above motivate a useful conceptual phase diagram [Fig.~\ref{fig:Fig1}(c)], constructed by extending models originally developed for disordered nearest-neighbor CSI systems~\cite{Savary2017,Benton2018,Pardini2019}. This description is formulated in the limit $|J_{\pm}|\ll J_1$, with $J_{\pm\pm}$ neglected, while allowing for finite, spatially varying $h_i$ that governs the phase behavior together with $J_1>0$. Within this framework, $\bar{h}$ denotes the spatial average of $h_i$, while $\delta h$ characterizes the width of its distribution, without assuming a specific functional form. 

Within this phase diagram, a QSI state is stabilized for weak and nearly uniform $h_i$, while two instabilities can drive the system out of this regime~\cite{Benton2018}. First, for a nearly uniform transverse field with $\bar{h} \gtrsim h_c \approx 0.6 J_1$, spinons condense, leading to a paramagnetic state composed of site-localized singlets~\cite{Roechner2016,Benton2018}. Second, when the width of the disorder distribution exceeds a critical threshold, $\delta h > \delta h_c$, spatial inhomogeneity in the ring-exchange amplitudes [Eq.~\ref{Hring}] localizes quantum tunneling to plaquettes proximate to disorder and promotes the local proliferation of visons in these regions~\cite{Savary2017,Benton2018}. This favors Ising ordering with a $(100)$ propagation vector~\cite{Chen2016MonopoleCondensation}; however, long-range order is suppressed by disorder, resulting instead in a spin-frozen state with frozen magnetic moments~\cite{Benton2018}. For QSI candidates, the critical disorder variance at vanishing average field, $\delta h_c$, is expected to remain finite due to quantum fluctuations arising from intrinsic transverse interactions. For finite $\delta h$~\cite{Griffiths1969,Savary2017}, Griffiths-type regimes may also emerge, in which rare regions locally satisfy $\bar{h} > h_c$ and behave paramagnetically, while the bulk remains either fluctuating (a “Griffiths QSI” regime) or frozen (a “Griffiths spin-frozen” regime).\par

\begin{figure*}[t] 
\centering
\includegraphics{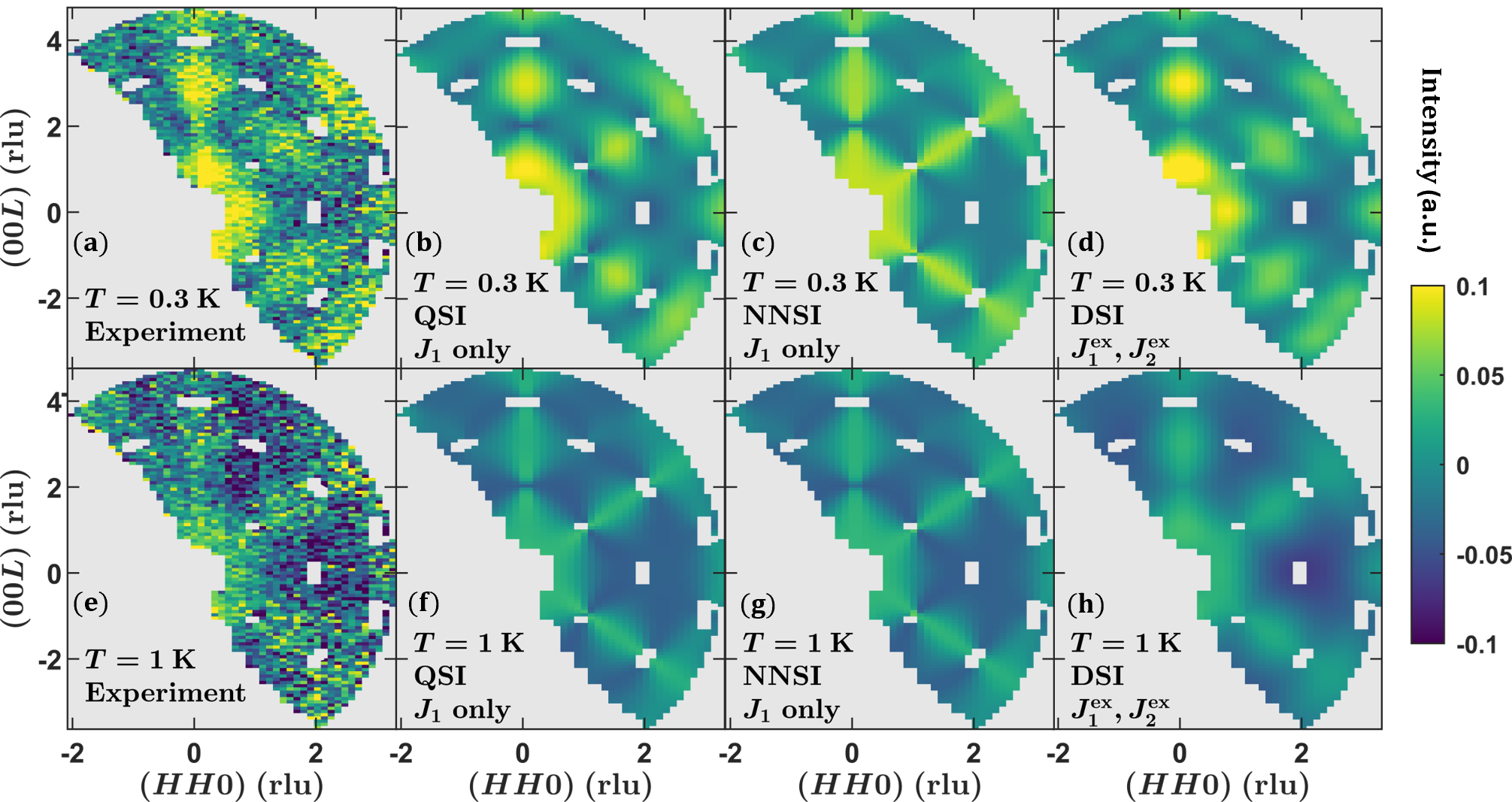}
\caption{\textbf{Magnetic diffuse scattering and comparison with spin-ice models.}
(a,e) Experimental magnetic diffuse scattering of \pso\ at $T = 0.3$~K and $T = 1$~K, respectively, after subtraction of high-temperature backgrounds ($T = 10$~K for (a) and $T = 20$~K for (e)). Data are integrated along $[K\bar{K}0]$ over $K \in [-0.1,\,0.1]$ reciprocal lattice units (rlu). 
(b--d,f--h) Corresponding simulations. Panels (b,c,f,g) show results for the spin-ice model of Ref.~\cite{Benton2012}: (b,f) in the quantum limit with a uniform ring-exchange amplitude $K^{\mathrm{eff}} = 0.0065(3)$~meV, and (c,g) for the classical nearest-neighbor spin-ice (NNSI) model. Panels (d,h) show best-fit results for the dipolar spin-ice (DSI) model including long-range dipolar interactions and further-neighbor exchange. Further details of the modeling are provided in the main text.}
\label{fig:Fig2}
\end{figure*}

A central challenge in realizing quantum spin liquids in real materials is understanding how weak disorder reshapes their emergent excitations and ground states, and drives systems into proximate phases within a disorder-influenced phase diagram. Among the Pr-based pyrochlores, \pso\ has long been considered a promising QSI candidate, hosting a non-Kramers ground-state doublet for Pr$^{3+}$ that is well separated from higher crystal-field levels~\cite{Princep2013Pr2Sn2O7_CrystalField}. Spins remain dynamically fluctuating down to at least 0.2~K without signatures of long-range order, as evidenced by low-energy magnetic scattering extending to $\sim 0.2$~meV~\cite{Zhou2008,Ortiz2024Pr2Sn2O7}.

In this work, we present the first comprehensive experimental study of flux-grown single-crystal \pso, combining neutron scattering with bulk thermodynamic measurements. We show that, at intermediate temperatures below $\sim 1$~K, the system exhibits clear signatures of quantum spin-ice behavior, including anisotropic spin-ice correlations and two well-separated dynamical timescales consistent with theoretical expectations and prior powder studies. Upon further cooling, the system undergoes a disorder-induced spin-freezing transition below $T_f \approx 0.15$~K, accompanied by a persistent gapped magnetic excitation and incipient $(100)$ correlations. These results place flux-grown \pso\ near the fully spin-frozen boundary in close proximity to the QSI phase and, when considered alongside prior studies, suggest a unified disorder-driven framework for interpreting the diverse behavior across Pr-based pyrochlores, with phenomenology reminiscent of Griffiths-type regimes.

\section{Results}
\subsection{Equal-time magnetic correlations}

Single-crystal magnetic diffuse scattering measured at $T = 0.3$~K and $T = 1$~K is shown in Fig.~\ref{fig:Fig2}a and Fig.~\ref{fig:Fig2}e, respectively. At $T = 1$~K [Fig.~\ref{fig:Fig2}e], \pso\ already exhibits a characteristic pinch-point–like pattern, signaling the onset of spin-ice correlations. Upon cooling to $T = 0.3$~K [Fig.~\ref{fig:Fig2}a], the scattering develops a more anisotropic and asymmetric structure. Compared with simulations of the classical nearest-neighbor spin-ice (NNSI) model~\cite{Benton2012} [Fig.~\ref{fig:Fig2}c], the experimental data display enhanced anisotropy along the $[001]$ and $[111]$ directions and broader intensity near the $(222)$ position, indicating that the $T = 0.3$~K data cannot be adequately described by the NNSI model alone.

To elucidate the origin of the observed $\boldsymbol{Q}$ dependence, we analyze the data using two complementary theoretical descriptions. First, we consider the QSI model of Ref.~\cite{Benton2012}, characterized by a uniform ring-exchange interaction $K^{\mathrm{eff}} \equiv K^{J} + \bar{K}^{h}$ [Eq.~\ref{Hring}], where $\bar{K}^{h}$ is taken as the average of $K^{h}_{ijklmn}$ over all hexagonal plaquettes. The best fit yields $K^{\mathrm{eff}} = 0.0065(3)$~meV. Second, we employ a classical dipolar spin-ice (DSI) model~\cite{paddison2023spinteract}, incorporating calculated dipolar interactions $D_{\mathrm{dip}} \approx 0.0010$~meV. Within this framework, the $n$th-neighbor coupling is given by $J_n = J_n^{\mathrm{ex}} + D_n$, where $J_n^{\mathrm{ex}}$ denotes the exchange contribution and $D_n$ the effective dipolar term (see Sec.\ref{APP:diffuse} of the Supplementary Materials for details). Satisfactory agreement requires exchange interactions beyond nearest neighbors, and can be achieved with $J_1^{\mathrm{ex}} = 0.0215(26)$~meV and second-neighbor exchange $J_2^{\mathrm{ex}} = -0.0016(1)$~meV. Notably, both the QSI model [Fig.~\ref{fig:Fig2}b,f] and the extended DSI model [Fig.~\ref{fig:Fig2}d,h] provide excellent descriptions of the measured scattering.

On the basis of equal-time correlations alone, it is therefore not possible to disentangle the roles of quantum dynamics and further-neighbor interactions. To resolve this ambiguity, we turn to the theoretical prediction that a QSI phase in \pso\ should exhibit two well-separated dynamical timescales~\cite{Tomasello2019}: a slow component, $\tau_{\mathrm{slow}}$, accessible via a.c.\ susceptibility, and a fast component, $\tau_{\mathrm{fast}}$, probed by inelastic neutron scattering. We therefore employ both techniques to investigate the spin dynamics.

\subsection{A.c. susceptibility}
\begin{figure*}[tp!] 
\centering
\includegraphics{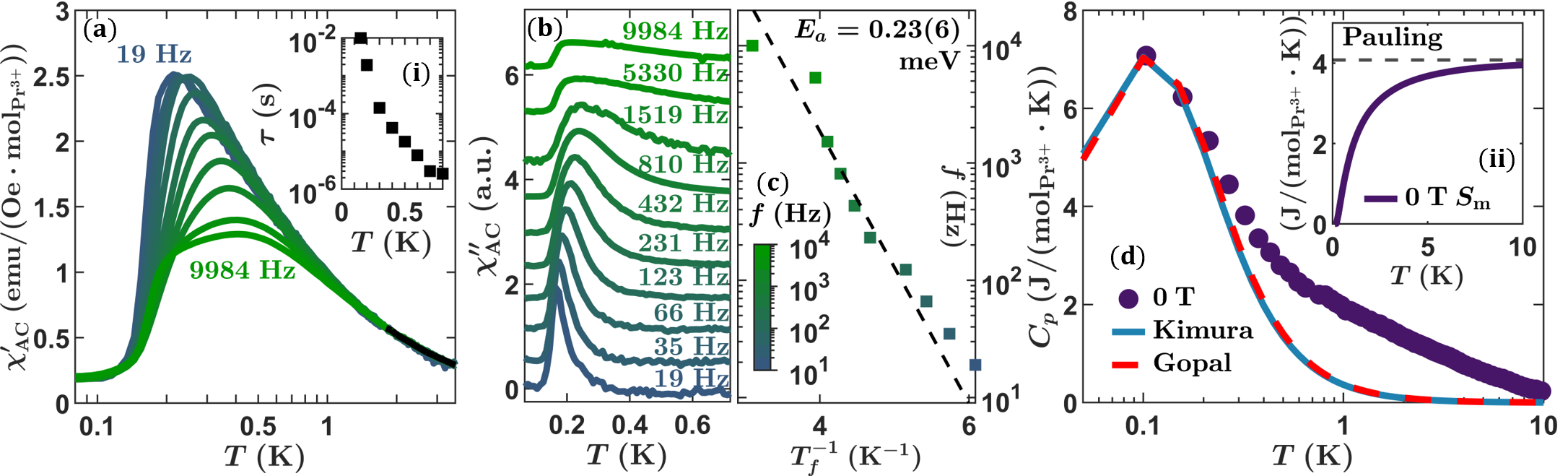}
\caption{\textbf{Bulk signatures of spin freezing in flux-grown single-crystal \pso.} 
(a) Temperature dependence of the real part of the zero-field a.c.\ susceptibility $\chi'$ measured along the $[111]$ direction at multiple frequencies  using an a.c.\ drive field of 2.5~Oe, normalized to the d.c.\ susceptibility in the range 1.8–4~K (black curve). Inset (i): Spin-relaxation time $\tau$ extracted from (a) using Eq.~\ref{dampHarm}. The extracted values should be regarded as order-of-magnitude estimates; error bars are omitted for clarity.
(b) Imaginary part $\chi''$ of the a.c.\ susceptibility measured under the same conditions. 
(c) Arrhenius fit of the frequency dependence of the $\chi''$ peak position, yielding an activation energy $E_a = 0.23(6)$~meV and an attempt frequency $f_0 \approx 80$~MHz, where $f_0$ should be regarded as an order-of-magnitude estimate. Frequencies used in (a,b) are indicated by the color bar. 
(d) Heat capacity as a function of temperature after subtraction of the lattice contribution estimated from non-magnetic La$_2$Sn$_2$O$_7$, compared with representative models of the nuclear Schottky anomaly. Inset (ii): Magnetic entropy obtained after subtracting the nuclear and lattice contributions (see Methods). The entropy released between the base temperature and 10~K is comparable to the Pauling value $R\!\left(\ln2-\tfrac{1}{2}\ln\tfrac{3}{2}\right)$. Owing to uncertainties in separating nuclear and electronic contributions, the entropy estimate is approximate (see Methods).
Error bars represent one standard deviation (s.d.); where not visible, they are smaller than the symbol size.}
\label{fig:Fig3}
\end{figure*}
As shown in Fig.~\ref{fig:Fig3}a,b, zero-field a.c.\ susceptibility measurements on single-crystal \pso\ reveal a progressive slowing of spin dynamics upon cooling below $\sim 1$~K. This is evidenced by the pronounced frequency dependence of the real component, $\chi'$, whose peak shifts to higher temperatures with increasing measurement frequency. Upon further cooling, the system undergoes a spin-freezing transition at $T_f \approx 0.15$~K, below which $\chi'$ is strongly suppressed and approaches zero. Consistently, the imaginary component $\chi''$ exhibits a clear peak at temperatures above $T_f$ for all measured frequencies, indicative of dissipative spin dynamics preceding the frozen state~\cite{Topping2019}.

Fitting $\chi'$ using a damped harmonic oscillator model [Eq.~\ref{dampHarm}]~\cite{Topping2019,Ortiz2024Pr2Sn2O7} yields a slow relaxation timescale $\tau_{\mathrm{slow}}$ spanning approximately $10^{-6}$–$10^{-2}$~s below 1~K, with a characteristic value $\tau_{\mathrm{slow}} \sim10^{-5}$-$10^{-4}$~s just above the freezing transition [Fig.~\ref{fig:Fig3}i]. The frequency dependence of $T_f$, determined from the peak position in $\chi''$ [Fig.~\ref{fig:Fig3}b], is well described by Arrhenius behavior, $f = f_0 \exp(-E_a/k_{\mathrm{B}}T_f)$, with an activation energy $E_a = 0.23(6)$~meV and an attempt frequency $f_0 \approx 80$~MHz, where $f_0$ should be regarded as an order-of-magnitude estimate [Fig.~\ref{fig:Fig3}c], in good agreement with previous results on powder samples~\cite{Ortiz2024Pr2Sn2O7}.

\subsection{Nuclear Schottky anomaly}

The spin-freezing transition is further reflected in the zero-field heat capacity, $C_p$ [Fig.~\ref{fig:Fig3}d], which displays a pronounced nuclear Schottky anomaly approaching the theoretical maximum value of $\sim 7~\mathrm{J}/(\mathrm{mol}_{\mathrm{Pr^{3+}}}\,\mathrm{K})$ at base temperature. This behavior is consistent with fully frozen Pr moments below $T_f$~\cite{Kimura2013,Tang2022_spinorbital}. After subtracting the hyperfine contribution~\cite{Kimura2013,Gopal1966}, the magnetic entropy released between the base temperature ($\sim0.1$~K) and 10~K is found to be comparable to the Pauling value $R\!\left(\ln2-\tfrac{1}{2}\ln\tfrac{3}{2}\right)$ [Fig.~\ref{fig:Fig3}ii]. Although a definitive separation of the nuclear and electronic contributions to the heat capacity is not possible, this result is broadly consistent with the expectation that most spin-ice entropy is released below 10~K and that spin-ice correlations are largely absent at 10~K.

\subsection{Elastic and inelastic neutron scattering}
\begin{figure*}[t] 
\centering
\includegraphics{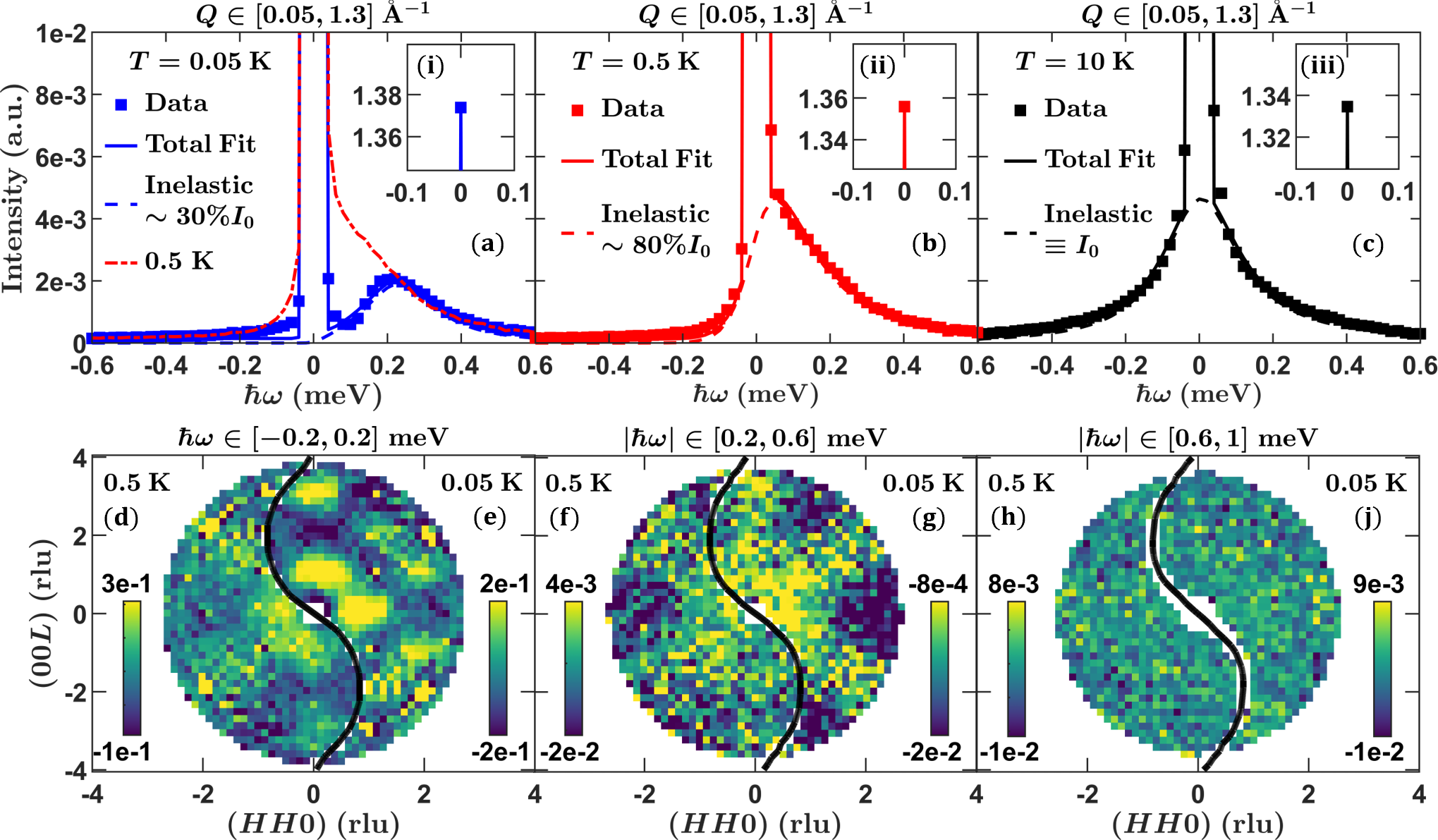}
\caption{\textbf{Temperature evolution of spin dynamics and correlations from neutron spectroscopy in single-crystal \pso.} 
(a--c) Powder-averaged scattering intensity as a function of energy transfer $\hbar\omega$ at $T = 0.05$, 0.5, and 10~K, measured with incident energy $E_i = 0.99$~meV. Data are averaged over $Q \in [0.05,\,1.3]~\mathrm{\AA}^{-1}$ and along $[K\bar{K}0]$ with $K \in [-0.1,\,0.1]$~rlu. For comparison, the $T = 0.5$~K spectrum is overlaid in (a). Insets (i--iii) show the corresponding elastic intensities. Dashed lines indicate fits to the dynamical response (see Methods). 
(d--j) $\boldsymbol{Q}$-dependent magnetic scattering (arb.\ units) at $T = 0.5$~K (d,f,h) and $T = 0.05$~K (e,g,j), measured with $E_i = 3.32$~meV after subtraction of the $T = 10$~K background. Data are integrated along $[K\bar{K}0]$ with $K \in [-0.15,\,0.15]$~rlu over energy windows $\hbar\omega \in [-0.2,\,0.2]$~meV (d,e), $[0.2,\,0.6]$~meV (f,g), and $[0.6,\,1.0]$~meV (h,j). No symmetrization has been applied. Energy-gain and energy-loss channels are combined in (f,g,h,j). Color scales are centered on the mean intensity of each panel. 
Error bars represent one standard deviation (s.d.); where not visible, they are smaller than the symbol size.}
\label{fig:Fig4}
\end{figure*}
\begin{figure*}[t] 
\centering
\includegraphics{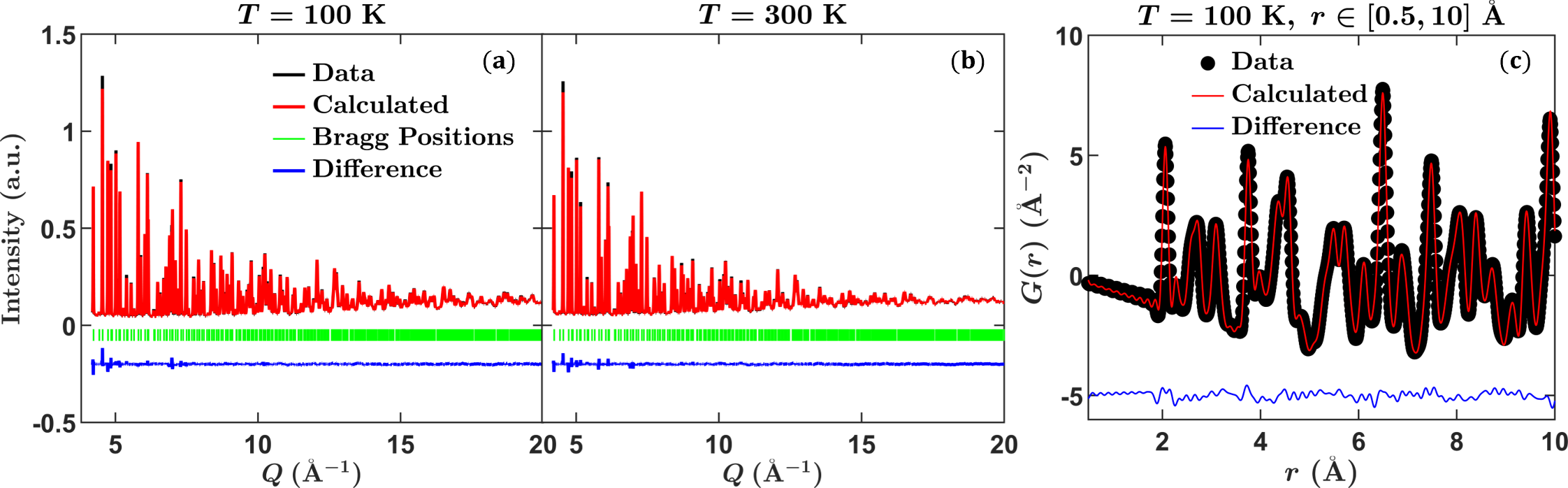}
\caption{\textbf{Neutron diffraction and pair distribution function analysis of pulverized flux-grown single-crystal \pso.} 
(a,b) Rietveld refinements of neutron powder diffraction patterns measured at $T = 100$~K and $300$~K. Observed data are shown as black symbols, calculated patterns as red curves, Bragg reflection positions as green tick marks, and the difference between observed and calculated intensities as blue curves. 
(c) Pair distribution function (PDF) analysis obtained from total neutron scattering data at $T = 100$~K. Experimental data (black), refined fit (red), and difference curve (blue) are shown. Error bars represent one standard deviation (s.d.); where not visible, they are smaller than the symbol size.}
\label{fig:Fig5}
\end{figure*}
Neutron spectroscopy provides key insight into spin dynamics both above and below $T_f$. The energy dependence of the powder-averaged scattering is shown in Fig.~\ref{fig:Fig4}a--c. At $T = 10$~K, pronounced inelastic scattering extends up to $0.6$~meV [Fig.~\ref{fig:Fig4}c], which we attribute to transitions between the two singlets of the non-Kramers ground-state doublet split by site-dependent transverse fields $h_i$~\cite{Wen2017,Martin2017}. These transitions are hereafter referred to as the \emph{singlet--singlet excitation}. The spectra are well described by a model of independent singlet--singlet excitations with energy transfer $\Delta_i = 2h_i$ [Eq.~\ref{Eq:10K_singlelevel}], assuming a half-Lorentzian distribution of $h_i$ [Eq.~\ref{Eq:halfLtz}]. The corresponding fit is shown as a dashed line in Fig.~\ref{fig:Fig4}c. The extracted half-width at half-maximum (HWHM) of the inelastic scattering is $\Gamma_{\Delta} = 2\Gamma_h = 0.13(1)$~meV, where $\Gamma_h = 0.063(3)$~meV is significantly smaller than the value $\Gamma_h = 0.27(1)$~meV reported for floating-zone--grown \pzo\ samples~\cite{Wen2017}.

Upon cooling below $T \sim 1$~K, spin-ice correlations progressively develop in \pso, leading to a pronounced evolution of the singlet--singlet excitation. The excitation shifts to higher energies, accompanied by a redistribution of spectral weight from the inelastic channel toward the quasielastic channel, where signatures of QSI physics may emerge. A mean-field description of this evolution is provided in Methods. At $T = 0.5$~K, both modified singlet--singlet excitations and finite-temperature QSI excitations are expected to contribute; however, these components cannot be resolved separately at the present experimental resolution. To phenomenologically describe the spectra, we model the response using a damped harmonic oscillator (DHO) form [Eq.~\ref{Eqn:DHO}], which yields overdamped dynamics with characteristic timescales on the order of $\tau \sim 10^{-11}$~s, comparable to the fast relaxation channel $\tau_{\mathrm{fast}}$ identified in Ref.~\cite{Tomasello2019}. Because a substantial fraction of the inelastic response originates from overlapping singlet--singlet excitations, the extracted timescale should be regarded as a lower bound on the intrinsic QSI relaxation time. The remaining low-energy spectral weight near zero energy transfer may arise from photon-like QSI excitations that remain unresolved from the elastic line.

At $T = 0.05$~K, well below $T_f$, the magnetic excitation spectrum undergoes a pronounced transformation. Relative to $T = 0.5$~K, the low-energy inelastic spectral weight in the range $[0,\,0.2]$~meV is strongly suppressed, while a well-defined gapped excitation persists. As QSI dynamics are expected to be suppressed below the spin-freezing transition, the inelastic response in this regime is attributed entirely to the modified singlet--singlet excitation. Analysis using the DHO form [Eq.~\ref{Eqn:DHO}] yields a damped propagating mode centered at $E_{\mathrm{gap}} = 0.23(1)$~meV with a HWHM $\Gamma = 0.12(2)$~meV. The gap energy is in good agreement with the activation energy $E_a = 0.23(6)$~meV extracted from a.c.\ susceptibility measurements. The linewidth $\Gamma$ is comparable to $\Gamma_{\Delta}$ at $T = 10$~K, indicating that the broadening is similarly governed by the distribution of quenched disorder.

By integrating the fitted dynamical response functions (dashed lines in Fig.~\ref{fig:Fig4}a--c), we find that the integrated intensity at $T = 0.05$~K and $T = 0.5$~K accounts for approximately $30\%$ and $80\%$, respectively, of the total fitted magnetic spectral weight observed at $T = 10$~K, denoted as $I_0$. Consistent with the total-moment sum rule, the spectral weight lost upon cooling is transferred to energies within the elastic resolution.

Two mechanisms likely contribute to the redistribution of spectral weight across the spin-freezing transition at $T_f$. First, the evolution from $T = 0.5$~K to $T = 0.05$~K qualitatively follows the expected temperature dependence of the singlet--singlet excitation as spin-ice correlations develop. However, this effect alone is unlikely to account for the full $\sim 50\%\,I_0$ transfer of spectral weight from the dynamical channel into the elastic channel, since spin-ice correlations are already well developed at $T = 0.5$~K [Fig.~\ref{fig:Fig4}d], and a substantial further enhancement of the exchange field at lower temperatures is not expected.

A second contribution arises from the suppression of QSI-related excitations present above $T_f$ upon entering the frozen state. In Pr-based QSI candidates, spin freezing has been theoretically predicted to favor $(100)$ magnetic order through vison condensation~\cite{Chen2016MonopoleCondensation,Benton2018}, consistent with the incipient $(100)$ correlations observed here (see below). A quantitative separation of these two contributions would require higher energy and temperature resolution than available in the present study.

We next examine the $\boldsymbol{Q}$ dependence of the magnetic scattering, shown in Fig.~\ref{fig:Fig4}d--j for $T = 0.05$~K and $T = 0.5$~K at representative energy ranges. These constant-energy slices are intended to highlight the momentum dependence of the magnetic scattering after background subtraction; quantitative comparisons of absolute intensity between panels are not meaningful (see Methods). At $T = 0.5$~K, the elastic scattering [Fig.~\ref{fig:Fig4}d] exhibits a pinch-point–like pattern, consistent with spin-ice correlations. Upon further cooling into the spin-frozen state at $T = 0.05$~K, the elastic scattering evolves markedly [Fig.~\ref{fig:Fig4}e], with intensity concentrated near reciprocal-space positions including $(001)$, $(003)$, $(110)$, $(1\pm2\,0)$, and $(2\pm1\,0)$. This redistribution suggests the development of short-range correlations with an incipient ordering wavevector $\boldsymbol{q} = (100)$ as the system freezes.

Similar tendencies have been reported in classical spin-ice (CSI) materials \hto~\cite{Bramwell2001,Kanada2002} and \dto~\cite{Yavorskii2008EmergentClusters,Samarakoon2022Glassiness} below their respective freezing temperatures. In those systems, however, deviations from ideal pinch-point behavior are weaker, and the freezing is commonly attributed to stronger dipolar interactions acting alongside geometric frustration~\cite{Samarakoon2022Glassiness}. By contrast, within the QSI framework, spin freezing of the magnetic dipole component corresponds to a confinement transition of the emergent $U(1)$ gauge field, which generically breaks translational symmetry. 
The simplest ordering tendency emerging from this theoretical framework is characterized by a propagation vector $\boldsymbol{q} = (100)$~\cite{Chen2016MonopoleCondensation,Benton2018}, qualitatively consistent with the diffuse scattering observed here.

The $\boldsymbol{Q}$ dependence of the gapped excitation at $T = 0.05$~K [Fig.~\ref{fig:Fig4}a] over the energy range $|\hbar\omega| \in [0.2,\,0.6]$~meV [Fig.~\ref{fig:Fig4}g] exhibits a characteristic ``star-fish-like'' pattern. A similar, albeit more diffuse, structure is observed at $T = 0.5$~K [Fig.~\ref{fig:Fig4}f] in the same energy range. By contrast, at higher energy transfers $\hbar\omega \gtrsim 0.6$~meV [Fig.~\ref{fig:Fig4}h,j], no significant $\boldsymbol{Q}$ dependence is resolved within the experimental sensitivity. These observations are consistent with singlet--singlet excitations modified by spin-ice correlations~\cite{Wen2017,Benton2018}, and closely resemble inelastic magnetic features reported in \pzo\cite{Kimura2013,Petit2016,Wen2017} and \pho~\cite{Sibille2018}. The contrasting $\boldsymbol{Q}$ dependence of the elastic and inelastic channels at $T = 0.5$~K further demonstrates that spin-ice correlations are well developed despite the presence of disorder-induced inelastic scattering.

\subsection{Characterization of structural disorder}

Our bulk measurements and neutron spectroscopy together indicate the presence of disorder in \pso, motivating a detailed examination of its structural origin. To this end, powder neutron diffraction measurements were performed on pulverized single-crystal material, and several candidate disorder models were tested via Rietveld refinement. These included: (i) partial oxidation of Pr, corresponding to Pr$_2$Sn$_2$O$_{7+y}$ with Pr$^{4+}$ occupying the Pr site; (ii) antisite cation disorder, in which Pr partially occupies the Sn site, modeled as Pr$_2$(Sn$_{2-x}$Pr$_x$)O$_{7+y}$; and (iii) Frenkel-type oxygen disorder, involving displacement of oxygen atoms to the interstitial $(\tfrac{1}{8},\,\tfrac{1}{8},\,\tfrac{1}{8})$ position of the cubic unit cell. In all cases, the best agreement with the diffraction data was obtained for models without resolvable disorder [Fig.~\ref{fig:Fig5}a,b], placing an upper bound of $\lesssim 1\%$ on the concentration of each disorder type considered. Consistently, no evidence for significant ionic off-centering was found, as the refined anisotropic atomic displacement parameters remain within typical values [Table~\ref{tab:refine}].

To probe disorder beyond the sensitivity of average-structure refinement, we performed neutron pair-distribution-function (PDF) analysis [Fig.~\ref{fig:Fig5}c], which provides a complementary real-space perspective. Refinement of the \textit{Fd$\bar{3}$m} pyrochlore structure using symmetry-allowed anisotropic atomic displacement parameters (ADPs) yields a significantly improved overall agreement with the data ($R_w = 7.71\%$) compared with an isotropic model ($R_w = 9.33\%$). The incorporation of anisotropic ADPs notably improves the description of the Pr--O and O--O correlations at interatomic distances of 2.587~\AA\ and 2.727~\AA, respectively, while slightly degrading the description of the nearest-neighbor metal--metal correlation at 3.746~\AA. These results indicate that the local disorder is predominantly accommodated through displacements of the oxygen sublattice, producing subtle distortions of the PrO$_8$ and/or SnO$_6$ coordination polyhedra while largely preserving the metal-ion positions defined by the average crystal structure. Together, these results demonstrate that disorder in \pso\ is weak and predominantly local, remaining essentially undetectable by conventional Rietveld refinement, yet still capable of strongly influencing the ultra-low-temperature magnetic behavior because of the non-Kramers nature of the system. In addition, the refined lattice parameter exhibits a weak dependence on the maximum real-space fitting range (10--20~\AA), consistent with the presence of local strains in the sample. The best-fit PDF refinement using anisotropic ADPs is shown in Fig.~\ref{fig:Fig5}c, while the refined structural parameters from both the Rietveld and PDF refinements are summarized in Table~\ref{tab:refine} of the Supplementary Materials.

\begin{table*}[ht]
\centering
\caption{\textbf{Comparison of experimental signatures in representative single-crystal (SC) and powder Pr-based pyrochlores.} 
For a.c.\ susceptibility, “partial freezing” denotes a finite $\chi'$ as $T \rightarrow 0$, whereas “complete freezing” indicates $\chi' \rightarrow 0$. The freezing temperature $T_f$ is determined from the peak in the lowest-frequency $\chi''$ as a function of temperature.}
\label{Tab:Pr_compare}
\begin{tabular}{lcccc}
\toprule
 & SC \pzo~\cite{Kimura2013} 
 & SC \pho~\cite{Sibille2016,Sibille2018} 
 & Powder \pso~\cite{Ortiz2024Pr2Sn2O7} 
 & SC \pso\ (this work) \\
\midrule

Preparation method 
 & Floating-zone 
 & Floating-zone 
 & Solid-state reaction 
 & Flux growth \\

\makecell[l]{Singlet--singlet \\ spectral weight (base $T$)}
 & \makecell{$\sim 90\%$ \\ ($T = 0.1$~K)}
 & \makecell{Subdominant (not quantified) \\ ($T = 0.05$~K)}
 & \makecell{$\sim 70\%$ \\ ($T \leq 0.5$~K)}
 & \makecell{$\sim 30\%$ \\ ($T = 0.05$~K)} \\

\makecell[l]{Nuclear Schottky peak \\
($\mathrm{J}\,(\mathrm{mol}_{\mathrm{Pr^{3+}}}\,\mathrm{K})^{-1}$)}
 & $\sim 2.6$ (reduced) 
 & Not reported 
 & $\sim 3.8$ (reduced) 
 & $\sim 7$ (full) \\

\makecell[l]{A.c.\ susceptibility $\chi'$ \\
as $T \rightarrow 0$}
 & Partial freezing 
 & \makecell{Tendency toward \\ complete freezing}
 & Partial freezing 
 & Complete freezing \\

Freezing temperature $T_f$ (K)
 & $\sim 0.15$
 & $< 0.075$
 & $\sim 0.25$
 & $\sim 0.15$ \\

\bottomrule
\end{tabular}
\end{table*}

\section{Discussion}

The convergence of multiple independent experimental probes demonstrates that our \pso\ samples exhibit magnetic properties broadly consistent with a QSI regime at finite temperatures below $\sim 1$~K. These include pronounced anisotropic spin-ice correlations together with two well-separated dynamical timescales, $\tau_{\mathrm{slow}} \sim10^{-5}$-$10^{-4}$~s and $\tau_{\mathrm{fast}} \sim 10^{-11}$~s, indicative of a bimodal distribution of magnetic dynamics associated with local spin-flip processes~\cite{Tomasello2019}. Upon further cooling, however, the system undergoes a complete spin-freezing transition at $T_f \approx 0.15$~K, signaling a clear departure from QSI behavior and the emergence of a distinct low-temperature phase as $T \rightarrow 0$.

This departure is most naturally attributed to weak structural disorder. Such an interpretation is supported by the pronounced sample dependence of ultra–low-temperature magnetic behavior reported for nominally identical compounds prepared via different synthesis routes. Although a quantitative characterization of disorder in \pso\ is beyond the scope of the present work, and the microscopic origin of disorder may vary across Pr-based pyrochlores~\cite{koohpayeh2014synthesis,Martin2017,Sibille2018,hicken2025intrinsic}, its impact on low-temperature magnetism can be understood within a unified framework based on Eq.~\ref{eq1} and the conceptual phase diagram shown in Fig.~\ref{fig:Fig1}c. To place our results within this framework, we compare $\Gamma_h = 0.063(3)$~meV (as a proxy for $\bar{h}$) with $J_1=J_1^{\mathrm{ex}} + D_1 \approx 0.033$~meV from the DSI analysis of magnetic diffuse scattering. These comparable energy scales ($\bar{h}/J_1 \sim \mathcal{O}(1)$) place our system near the regime where uniform-field theory~\cite{Roechner2016,Benton2018} predicts an instability of the QSI phase at $h_c \approx 0.6J_1$, highlighting the role of transverse fields in destabilizing QSI behavior.

In this regime, both disorder-induced states proximate to the QSI phase are relevant for interpreting our results. The first is a paramagnetic regime, characterized by a broad inelastic continuum of singlet--singlet excitations that remains dominant down to base temperature, despite a partial reduction of spectral weight associated with the development of spin-ice correlations, which also impart a ``star-fish-like'' $\boldsymbol{Q}$ dependence to the excitations~\cite{Wen2017,Martin2017,Benton2018}. The second is a spin-frozen regime, signaled by slow dynamics in a.c.\ susceptibility, the emergence of a nuclear Schottky anomaly in low-temperature heat capacity, and incipient magnetic correlations consistent with vison condensation~\cite{Chen2016MonopoleCondensation}. 

A comparison of these experimental diagnostics across representative Pr-based pyrochlores is summarized in Table~\ref{Tab:Pr_compare}. Within the framework of disorder-perturbed non-Kramers systems, these two instabilities are not expected to be mutually exclusive. Instead, the phenomenology exhibited by the materials listed in Table~\ref{Tab:Pr_compare} is broadly consistent with a Griffiths-type spin-frozen framework, interpolating between paramagnetic and frozen limits in which signatures of both may coexist. Here, we use “Griffiths spin-frozen state” in a phenomenological sense to describe such coexistence, without requiring spatially rare paramagnetic regions in the strict microscopic sense~\cite{Griffiths1969}.

Single-crystal \pzo~\cite{Kimura2013} and powder \pso~\cite{Ortiz2024Pr2Sn2O7} exhibit behavior that lies closer to the paramagnetic side of this regime. In both cases, the nuclear Schottky anomaly is substantially reduced from its theoretical maximum of $\sim 7~\mathrm{J}/(\mathrm{mol}_{\mathrm{Pr^{3+}}}\,\mathrm{K})$~\cite{Kimura2013}, reflecting suppression of the electronic moment by disorder-induced splitting of the non-Kramers doublet~\cite{Martin2017}. Consistently, a.c.\ susceptibility measurements reveal only partial spin freezing, with $\chi'$ remaining finite as $T \rightarrow 0$, while neutron spectra at the lowest reported temperatures remain dominated by inelastic singlet--singlet excitations. For powder \pso\ in Ref.~\cite{Ortiz2024Pr2Sn2O7}, neutron spectroscopy may probe an effective sample temperature ($\lesssim 0.5$~K) higher than the nominal base temperature ($0.02$~K) due to thermalization limitations. A definitive determination of the ground-state behavior would therefore benefit from future measurements under well-controlled thermalization conditions that access temperatures at or below the freezing scale ($\sim 0.25$~K).

By contrast, the flux-grown \pso\ crystals studied here exhibit experimental signatures that place them more clearly on the spin-frozen side of the phase diagram. These include a strongly reduced inelastic singlet--singlet spectral weight at base temperature, recovery of the full nuclear Schottky anomaly ($\sim 7~\mathrm{J}/(\mathrm{mol}_{\mathrm{Pr^{3+}}}\,\mathrm{K})$), a near-complete suppression of $\chi'$ as $T \rightarrow 0$, and incipient $(100)$ correlations observed in neutron scattering. Floating-zone–grown \pho~\cite{Sibille2016,Sibille2018} displays several qualitatively similar trends, including subdominant inelastic singlet--singlet spectral weight~\cite{Sibille2018} and a pronounced decrease of $\chi'$ toward zero upon cooling below $\sim 0.1$~K~\cite{Sibille2016}, indicative of proximity to a spin-frozen regime. However, in \pho\ the freezing temperature appears to lie at or below the lowest experimentally accessible temperatures ($\lesssim 0.075$~K)~\cite{Sibille2016}, and elastic neutron-scattering measurements at $T = 0.05$~K do not reveal clear incipient magnetic order~\cite{Sibille2018}. Taken together, these observations suggest that \pho\ resides near the boundary of a spin-frozen regime comparable to that realized in flux-grown single-crystal \pso, with the differences most naturally attributed to a lower freezing temperature and sample-dependent disorder, rather than any qualitative distinction in the underlying physics.

More generally, these observations suggest that the diverse behavior of Pr-based pyrochlores may be understood in terms of the strength and spatial distribution of disorder-induced transverse fields arising from quenched disorder and local structural distortions. Across different compounds, the relative strength of transverse fields and magnetic interactions, parameterized by $\bar{h}/J_1$, appears to govern the extent to which collective spin-ice correlations survive in the presence of disorder. Within the disorder-dominated regime, the spatial distribution of transverse fields, characterized by $\delta h$, may further influence whether the resulting state is predominantly paramagnetic or spin-frozen. Relatively uniform transverse fields are expected to favor a singlet-dominated paramagnetic regime, whereas strongly inhomogeneous transverse fields can promote coexistence of weakly and strongly perturbed regions, leading to spin-frozen behavior.

Our \pso\ samples therefore realize a disorder-induced phase proximate to the ideal $U(1)$ quantum spin liquid, exhibiting coexistence of frozen and paramagnetic characteristics. This phenomenology bears similarities to Griffiths-type behavior discussed in previous theoretical work on disorder-perturbed non-Kramers pyrochlores~\cite{Savary2017,Benton2018}. A contrasting limiting scenario is a more homogeneous frozen state, in which the average transverse field $\bar{h}$ lies just below the critical value $h_c$. In practice, these two limits are difficult to distinguish using bulk measurements, and both are broadly consistent with the observed coexistence of elastic and inelastic magnetic scattering. Although the present measurements do not directly establish the existence of spatially rare paramagnetic regions embedded within an otherwise frozen magnetic matrix, the absence of resolvable disorder in Rietveld refinement, despite signatures of local disorder from PDF analysis, is suggestive of a Griffiths-like scenario in which sparse defects generate locally enhanced transverse fields that strongly perturb nearby Pr$^{3+}$ environments. Definitive discrimination between Griffiths-type and more homogeneous disorder-induced frozen states would require local probes, such as muon spin relaxation ($\mu$SR), which are sensitive to spatial variations of internal magnetic fields. More broadly, our results demonstrate how weak structural disorder can qualitatively reshape the low-temperature behavior of candidate quantum spin liquids, stabilizing frozen phases that nevertheless retain clear dynamical and spatial signatures of their proximate QSI parent state.

\section{Methods}

\subsection{Sample preparation}

Single crystals of \pso\ were grown via a flux method using a 1:1.2 mass ratio of dried Na$_2$B$_4$O$_7$:NaF as the flux. This mixture was combined with Pr$_6$O$_{11}$, SnO$_2$, and SnO to yield a nominal 1:5 mass ratio of Pr$_2$Sn$_2$O$_7$ to flux. The resulting mixture was thoroughly ground and loaded into a 20~mL platinum crucible with a loosely fitting platinum lid, heated to $1000^\circ$C at $200^\circ$C/h, held for 6~h, and then cooled to $500^\circ$C at $1$–$1.5^\circ$C/h. Approximately 50\% of the flux mass was lost due to volatilization. Residual flux was removed by dissolution in boiling water. The resulting crystals are well-faceted, dark-red octahedra, typically $1$–$20$~mg in mass and millimeter-sized.

For single-crystal neutron scattering experiments, a total of 175 well-faceted octahedral crystals were mounted on both sides of two oxygen-free copper plates using GE varnish. The crystals covered a rectangular area of approximately $1.5$~cm (vertical) $\times$ $1.0$~cm (horizontal) on each plate, with the $(HHL)$ scattering plane oriented horizontally. The resulting assembly behaves effectively as a single crystal with a modest mosaic spread, with a total mass of approximately $1.27$~g and an estimated mosaicity of $\sim 5^\circ$ full width at half maximum (FWHM).

For neutron powder diffraction measurements performed on the Nanoscale-Ordered Materials Diffractometer (NOMAD), approximately 500~mg of single-crystal material was gently crushed and sealed in a 3-mm-diameter quartz capillary. The resulting fine-grained powder (or microcrystalline aggregate) exhibits a characteristic dark-red coloration. Photographs of the samples, along with diffraction measurements confirming the mosaicity, are provided in Fig.~\ref{fig:SIFig1} of the Supplementary Materials.

\subsection{Bulk characterization}

A single-crystal specimen with a mass of approximately 1~mg was used for heat capacity and a.c.\ susceptibility measurements using a Quantum Design 9~T Dynacool Physical Property Measurement System (PPMS), covering both the conventional $^4$He and dilution refrigerator (DR) temperature ranges. Additional bulk characterization, including magnetization-versus-field measurements and d.c.\ susceptibility data, is provided in Sec.~\ref{APP:sample} and Fig.~\ref{fig:SIFig4} of the Supplementary Materials.

For the heat capacity data, the nuclear Schottky anomaly in $C_p$ was analyzed using two models: (1) the analytical model of Kimura \textit{et al.}~\cite{Kimura2013}, with $\mu_{\mathrm{hyp}}^{\mathrm{Pr}} = 1.9~\mu_{\mathrm{B}}$ and no scaling factor; and (2) a statistical (Gopal) model~\cite{Gopal1966}, which treats the system as a multilevel scheme with six evenly spaced energy levels (nuclear spin $I = \frac{5}{2}$ for $^{141}\mathrm{Pr}$), with a level spacing of $\epsilon_r = 0.011$~meV. Both models reproduce the broad maximum in $C_p$ near 0.1~K. The heat capacity data and parameters obtained from model (1) closely resemble those reported for \pzo\ sample 1 in Ref.~\cite{Tang2022_spinorbital}. 

The magnetic entropy $S_m$ [Fig.~\ref{fig:Fig3}(ii)] was obtained by subtracting both the nuclear contribution (modeled using the Gopal approach) and the lattice contribution (estimated from non-magnetic La$_2$Sn$_2$O$_7$) from the measured $C_p$, and integrating the resulting magnetic contribution $C_m/T$. Subtraction using either the Kimura or Gopal model yields essentially identical results for the magnetic entropy. We note that a definitive separation between nuclear Schottky and electronic magnetic contributions to the heat capacity is not possible. Consequently, the fitted nuclear Schottky peak may be slightly overestimated, with a corresponding excess in the measured heat capacity arising from electronic magnetic contributions. The magnetic entropy shown in Fig.~\ref{fig:Fig3}(ii) should therefore be regarded as an approximate lower-bound estimate. In particular, the entropy recovered between the base temperature ($\sim0.1$~K) and 10~K may be somewhat larger than the value close to $R\!\left(\ln2-\tfrac{1}{2}\ln\tfrac{3}{2}\right)$ obtained from the present subtraction procedure.

To extract the ``slow'' timescale of spin dynamics~\cite{Tomasello2019} from the real part $\chi'$ of the a.c.\ susceptibility, we employ a damped harmonic oscillator model~\cite{Topping2019,Ortiz2024Pr2Sn2O7}:
\begin{equation}
\chi'(f) = \chi_S + \frac{\chi_T - \chi_S}{1 + (2\pi f \tau)^2}, \label{dampHarm}
\end{equation}
where $f$ is the driving frequency, $\tau$ is the characteristic spin correlation time, $\chi_S$ is the adiabatic (high-frequency) susceptibility, and $\chi_T$ is the isothermal (low-frequency) susceptibility. Details of the fits at each temperature are provided in Fig.~\ref{fig:SIFig2} of the Supplementary Materials.

\subsection{Single-crystal neutron scattering experiment}

Magnetic diffuse scattering measurements were performed on the Wide-Angle Neutron Diffractometer (WAND$^2$) at the High Flux Isotope Reactor (HFIR), Oak Ridge National Laboratory (ORNL), using unpolarized neutrons with an incident wavelength $\lambda = 1.5~\text{\AA}$. Two primary data sets were collected at $T = 0.3$~K and $T = 1$~K. Background subtraction was performed using high-temperature measurements at $T = 10$~K and $T = 20$~K, respectively. The use of two background data sets compensates for a minor change in background conditions during the experiment. As indicated by the heat-capacity analysis, most spin-ice entropy is released below 10 K (see Fig.~\ref{fig:Fig3}ii). Spin-ice correlations are therefore expected to be largely absent for $T \geq 10$~K, and the magnetic structure factor should consequently exhibit only weak $\boldsymbol{Q}$ dependence ($\propto |F(Q)|^2$, where $F(Q)$ is the magnetic form factor of Pr$^{3+}$), making the $T\geq10$ K data an appropriate reference for removing the nuclear background and isolating the $\boldsymbol{Q}$ dependence of the magnetic scattering. Inelastic neutron scattering measurements were carried out on the Cold Neutron Chopper Spectrometer (CNCS) at the Spallation Neutron Source (SNS), ORNL. Data were collected at $T = 0.05$~K, $0.5$~K, and $10$~K using incident neutron energies of $0.99$ and $3.32$~meV. High-flux chopper configurations were employed to optimize energy resolution and reciprocal-space coverage, yielding elastic energy resolutions (FWHM) of approximately $0.02$ and $0.1$~meV, respectively.

The data shown in Fig.~\ref{fig:Fig4}a--c were processed to remove a weak background feature present at all temperatures; details are provided in Sec.~\ref{APP:sub} (see also Figs.~\ref{fig:SIFig5} and \ref{fig:SIFig6}) of the Supplementary Materials. For the slices in Fig.~\ref{fig:Fig4}d--j, data collected at $T = 10$~K were subtracted to isolate the magnetic scattering, following the same rationale as for the WAND$^2$ data. For the constant-energy slices shown in Fig.~\ref{fig:Fig4}d--j, the color scale of each panel is centered on its mean intensity after subtraction of the corresponding $T=10$ K background. For a given energy range, the color-scale span is chosen to be identical for the $T=0.05$ K and $T=0.5$ K datasets. Consequently, these plots are designed to facilitate comparison of the momentum dependence of the scattering between temperatures within each energy transfer, rather than quantitative comparison of absolute intensities across different energy transfers. The energy integration windows were chosen to avoid leakage from the elastic line; the corresponding resolution analysis is presented in Sec.~\ref{APP:CNCS_ConstE} (see also Fig.~\ref{fig:SIFig9}). Energy-gain and energy-loss channels were combined to improve statistical accuracy and to account for temperature-dependent background variations and possible non-equilibrium effects associated with spin freezing. More details and the separated channels are presented in Sec.~\ref{APP:CNCS_ConstE_Sep} and Fig.~\ref{fig:SIFig10}. Additional CNCS data are provided in Figs.~\ref{fig:SIFig5}, \ref{fig:SIFig7}, and \ref{fig:SIFig8} of the Supplementary Materials.

For both WAND$^2$ and CNCS measurements, the sample was rotated about the vertical $[1\bar{1}0]$ axis over a total angular range of $180^\circ$, with step sizes of $0.1^\circ$ and $1^\circ$, respectively. All neutron-scattering data were reduced and processed using the Mantid software suite~\cite{Arnold_2014_Mantid}.

\subsection{Mean-field model of split non-Kramers doublets}

To interpret the temperature evolution of the singlet--singlet excitation, we summarize a local mean-field framework developed for disordered non-Kramers pyrochlores~\cite{Martin2017,aczel2026influence}. Each Pr$^{3+}$ site $i$ can be described by an effective two-level Hamiltonian,
\begin{equation}
\mathcal{H}_{\mathrm{mf}} =
\begin{pmatrix}
0 & -h_i \\
-h_i & E_J
\end{pmatrix},
\label{HMF}
\end{equation}
expressed in the basis of pseudospin states aligned parallel ($\ket{+}$) and antiparallel ($\ket{-}$) to the local $\hat{\boldsymbol{z}} \parallel \langle 111\rangle$ easy axis. The parameter $h_i$ is defined in Eq.~\ref{eq1}, while $E_J$ represents the longitudinal spin-flip energy cost associated with transitions between $\ket{+}$ and $\ket{-}$. 

Within a mean-field approximation, $E_J$ is determined by the surrounding pseudospin configuration,
\begin{equation}
E_J = 2\sum_{n,j} J_n \langle \sigma_j^z \rangle,
\end{equation}
where the sum runs over the $n$th-nearest neighbors $j$ of site $i$. Retaining only nearest-neighbor interactions $J_1$ and assuming a local two-in–two-out configuration yields $E_J = 4J_1$. For clarity, transverse exchange terms $J_{\pm}$ and $J_{\pm\pm}$ in Eq.~\ref{eq1} are neglected. At the mean-field level, inclusion of $J_{\pm}$ would primarily renormalize the effective transverse field via $h_i \rightarrow h_i + J_{\pm}\sum_{n,j}\langle \sigma_j^x \rangle$ once quadrupolar correlations develop at low temperature.

The Hamiltonian in Eq.~\ref{HMF} yields a split spectrum consisting of a ground state $\ket{\mathrm{GS}}$ and an excited state $\ket{\mathrm{EX}}$, separated by an energy gap $\Delta = \sqrt{E_J^2 + 4h_i^2}$. The elastic and inelastic neutron-scattering intensities associated with transitions within the split doublet are
\begin{align}
\begin{split}
I_{\mathrm{el}} &\propto \left| \bra{\mathrm{GS}} \sigma^z \ket{\mathrm{GS}} \right|^2 = \frac{E_J^2}{\Delta^2}, \\
I_{\mathrm{inel}}(\Delta) &\propto \left| \bra{\mathrm{EX}} \sigma^z \ket{\mathrm{GS}} \right|^2 = \frac{4h_i^2}{\Delta^2},
\end{split}
\label{Iel_inel}
\end{align}
since neutrons predominantly probe the magnetic dipole moment along the local easy axis, and $\hat{J}^z \propto \sigma^z$~\cite{Petit2016,Chen2017DiracMonopoles} within the effective two-level subspace.

In the spin-ice limit ($h_i = 0$), the pseudospins align along $\hat{\boldsymbol{z}}$. The ground state $\ket{\mathrm{GS}}$ corresponds to the “two-in, two-out” configuration, while $\ket{\mathrm{EX}}$ corresponds to a single spin-flip defect (i.e., the creation of a pair of spinons). In this limit, the inelastic matrix element vanishes ($I_{\mathrm{inel}} = 0$), and the spectral weight resides entirely in the elastic channel. In the opposite limit ($h_i \gg E_J$), the pseudospin is rotated nearly perpendicular to the easy axis, yielding a singlet ground state $\ket{\mathrm{GS}} \rightarrow (\ket{+}+\ket{-})/\sqrt{2}$. Consequently, the elastic contribution is suppressed ($I_{\mathrm{el}} \rightarrow 0$), and the spectral weight is transferred entirely to the inelastic channel at energy $\Delta$.

At high temperatures, spin-ice correlations are negligible and the exchange field vanishes to leading order ($E_J \approx 0$). In this regime, the low-energy magnetic response is governed by singlet--singlet excitations with local splittings $\Delta_i = 2h_i$, distributed according to a probability density $p(\Delta)$. Upon cooling below $T \sim 1$~K, spin-ice correlations progressively develop, generating a finite exchange field $\sum_{n,j} J_n \langle \sigma_j^z \rangle$ (and hence $E_J$). Within this mean-field picture, the excitation energies shift to higher values, $\Delta_i \rightarrow \sqrt{(2h_i)^2 + E_J^2}$. Simultaneously, the growing exchange field favors alignment of the pseudospins along the local $\hat{\boldsymbol{z}}$ axis, leading to a progressive transfer of spectral weight from the inelastic to the elastic channel for sites where $h_i \sim E_J$.

\subsection{Fitting neutron spectroscopy data}
The spectra in Fig.~\ref{fig:Fig4}a--c were analyzed using the composite fitting function
\begin{equation}
\begin{split}
I(\hbar\omega) &= a_0 \exp\!\left[-\frac{(\hbar\omega)^2}{2\sigma^2}\right]
+ \int d\omega' \, R(\omega-\omega')\, S(\omega') + C_0,
\label{Inel_fit}
\end{split}
\end{equation}
where the Gaussian term accounts for elastic contributions, including nuclear scattering, static magnetic correlations, and residual instrumental background, as illustrated in Fig.~\ref{fig:Fig4}i--iii. The Gaussian width $\sigma \approx 0.0087$~meV was fixed to the instrumental energy resolution. The second term represents the quasielastic and inelastic magnetic response $S(\omega)$ convoluted with the instrumental energy resolution function,
\begin{equation}
R(\omega) = \frac{1}{\sqrt{2\pi}\sigma}\exp\!\left[-\frac{(\hbar\omega)^2}{2\sigma^2}\right],
\end{equation}
where the same value of $\sigma$ was used. The constant term $C_0$ accounts for a small residual background.

At $T = 10$~K, where spin-ice correlations are negligible, the system can be treated as an ensemble of effectively independent doublets, each split by a local energy $\Delta_i = 2h_i$ at site $i$. Starting from generalized susceptibility for a two-level system split by energy $\Delta$, we have~\cite{Martin2017}
\begin{equation}
\begin{split}
\chi(\omega,\Delta)
&=|\zeta(\Delta)|^2
\tanh\!\left(\frac{\Delta}{2 k_{\mathrm{B}} T}\right)
\left[
\frac{1}{\hbar\omega - \Delta + i0^+}\right.\\&
\left.-
\frac{1}{\hbar\omega + \Delta + i0^+}
\right].
\end{split}
\end{equation}
Here $\zeta(\Delta)=\bra{\mathrm{EX}}\sigma^z\ket{\mathrm{GS}}$ is the matrix element of dipole moment operator between the ground state $\ket{\mathrm{GS}}$ and the excited state $\ket{\mathrm{EX}}$ of the split doublet, and we have $\zeta(\Delta)\sim1$ at $T=10$ K. This leads to spin-spin correlation function for a two-level system $S(\omega,\Delta)$ through the fluctuation–dissipation theorem:
\begin{equation}
\begin{split}
S(\omega,\Delta)&=\frac{1}{\pi}\frac{1}{1-\exp(-\hbar\omega/k_{\mathrm{B}}T)}\chi''(\omega,\Delta)\\
&=\frac{|\zeta(\Delta)|^2}{\pi}\frac{1}{1-\exp(-\hbar\omega/k_{\mathrm{B}}T)}\tanh\!\left(\frac{\Delta}{2 k_{\mathrm{B}} T}\right)\\&\times\left(\delta(\hbar\omega-\Delta)-\delta(\hbar\omega+\Delta)\right)\label{Eq:10K_singlelevel}
\end{split}
\end{equation}

Following Ref.~\cite{Wen2017}, we model the effects of quenched disorder by assuming a half-Lorentzian distribution of the transverse field $h_i$,
\begin{equation}
p(h) = \frac{2}{\pi}\frac{\Gamma_h}{h^2 + \Gamma_h^2}, \qquad h \geq 0,\label{Eq:halfLtz}
\end{equation}
with $p(h) = 0$ for $h < 0$. This gives a distribution of local split energy $\Delta$,
\begin{equation}
p(\Delta) = \frac{2}{\pi}\frac{2\Gamma_h}{\Delta^2 + (2\Gamma_h)^2}, \qquad \Delta \geq 0,
\end{equation}
with $p(\Delta) = 0$ for $\Delta < 0$. The dynamical correlation function averaged over this distribution is then given by
\begin{equation}
\begin{split}
S(\omega) &= \int_0^{\infty} \mathrm{d}\Delta \, p(\Delta)\, S(\omega,\Delta) \\
&= C \, \frac{\tanh\!\left(\frac{\hbar\omega}{2k_{\mathrm{B}} T}\right)}{1 - \exp(-\hbar\omega / k_{\mathrm{B}} T)} 
\frac{2\Gamma_h}{(\hbar\omega)^2 + (2\Gamma_h)^2} \\
&= \frac{C}{1 + \exp(-\hbar\omega / k_{\mathrm{B}} T)} 
\frac{2\Gamma_h}{(\hbar\omega)^2 + (2\Gamma_h)^2},
\end{split}\label{Eqn:10K}
\end{equation}
where $C$ is an overall intensity scale factor. This expression is used to fit the spectra at $T = 10$~K [Fig.~\ref{fig:Fig4}c], yielding $\Gamma_h = 0.063(3)$~meV. For comparison with lower temperatures, the total magnetic scattering intensity $I_0$ is defined by integrating Eq.~\ref{Eqn:10K} over $\hbar\omega \in [-0.6,\,0.6]$~meV.

To phenomenologically describe the inelastic response below $T \sim 1$~K, we adopt a damped harmonic oscillator (DHO) form~\cite{petit2012spin},
\begin{equation}
\begin{split}
S(\omega) &= C_1
\frac{\hbar\omega}{1 - \exp(-\hbar\omega / k_{\mathrm{B}} T)}
\frac{1}{\left[(\hbar\omega)^2 - f^2\right]^2 + (\hbar\omega z)^2},
\end{split}
\label{Eqn:DHO}
\end{equation}
where $f$ denotes the characteristic mode frequency and $z$ is the damping parameter, both expressed in energy units. $C_1$ is an overall intensity scale factor.

At $T = 0.05$~K [Fig.\ref{fig:Fig4}a], fitting the spectra with Eq.~\ref{Eqn:DHO} yields $f = 0.26(1)$~meV and $z = 0.24(3)$~meV, corresponding to the underdamped regime ($f > z/2$)~\cite{lamsal2016extracting}. In this limit, the response is well approximated by a double-Lorentzian form,
\begin{equation}
\begin{split}
S(\omega) &\sim \frac{C_1}{1 - \exp(-\hbar\omega / k_{\mathrm{B}} T)} \\
&\quad \times \left[
\frac{\Gamma}{(\hbar\omega - E_{\mathrm{gap}})^2 + \Gamma^2}
- \frac{\Gamma}{(\hbar\omega + E_{\mathrm{gap}})^2 + \Gamma^2}
\right],
\end{split}
\label{Eqn:DL}
\end{equation}
with $E_{\mathrm{gap}} = \sqrt{f^2 - z^2/4} = 0.23(1)$~meV and $\Gamma = z/2 = 0.12(2)$~meV. The linewidth ($\Gamma$) is comparable to $\Gamma_{\Delta}=2\Gamma_h=0.13(1)$~meV obtained at $T = 10$~K, indicating that it is primarily governed by the distribution of quenched disorder. Integration over $\hbar\omega \in [-0.6,\,0.6]$~meV shows that the gapped excitation accounts for approximately $30\%$ of $I_0$.

At $T = 0.5$~K [Fig.\ref{fig:Fig4}b], fitting the spectra with Eq.~\ref{Eqn:DHO} yields $f = 0.17(1)$~meV and $z = 0.39(3)$~meV, placing the system in the overdamped regime ($f < z/2$)~\cite{lamsal2016extracting}. The response can then be interpreted as two overdamped relaxational modes with damping rates (expressed in energy units) $\Gamma_{1,2} = z/2 \pm \sqrt{z^2/4 - f^2} = 0.29(5)$~meV and $0.10(2)$~meV, corresponding to timescales $\tau_{1,2} = \hbar/\Gamma_{1,2} = 2.3(4)$ ps and $6.6(1.6)$ ps. Integrating Eq.~\ref{Eqn:DHO} over the energy range $\hbar\omega \in [-0.6,\,0.6]$~meV shows that the low-energy dynamical response, which is predominantly quasielastic within the present energy resolution, accounts for approximately $80\%$ of $I_0$. We note that such quasielastic spectral weight may arise either from intrinsic relaxational dynamics or from unresolved low-energy inelastic excitations, such as photon-like modes expected in a QSI regime.

\subsection{Powder neutron refinement}
Powder neutron-diffraction data were collected from pulverized single-crystal specimens at $T = 100$~K and $T = 300$~K. Rietveld refinements of the diffraction patterns were carried out using the software package \textsc{TOPAS}~\cite{Coelho2018TOPAS}. Neutron pair distribution function (PDF) analysis was performed using the \textsc{PDFgui}~\cite{farrow2007pdffit2} software package. For PDF analysis, a stoichiometric structural model was refined over short real-space distances up to 10~\AA. 

\paragraph{Uncertainty estimation.}
Unless otherwise stated, uncertainties of data points in all figures correspond to one standard deviation (1$\sigma$) statistical errors. For parameters obtained from least-squares or $\chi^2$ minimization procedures in diffuse-scattering fits, uncertainties were determined from the parameter range satisfying $\chi^2 < \chi^2_{\min}\left(1+\frac{1}{N-M}\right)$, where $N$ is the number of data points and $M$ is the number of fitted parameters, taking into account the numerical resolution of the parameter grid. For all other fitted quantities, uncertainties were estimated from covariance-matrix analysis near the best-fit solution. Systematic uncertainties are not included.

\section{Data Availability}
All the data supporting the findings of this study are available within the article and from the corresponding authors upon request. A Processed Data file containing the numerical data underlying all plotted figures is available at \href{https://docs.google.com/spreadsheets/d/e/2PACX-1vSnMLHU11eZI3Ya54i2F2onRnBWZzTT7ZL4owD5nyPBSGBzQXt9rCoZ0vk2kyMjHD6r_wZ3FY_wi0KA/pubhtml}{Processed Data}.

\section{Acknowledgments}
We thank Owen Benton and Gang Chen for insightful theoretical discussions, and Danielle R.~Yahne for valuable discussions on the structural refinement. B.R.O. (bulk property measurements and analysis, crystallographic studies) gratefully acknowledges support from the U.S. Department of Energy (DOE), Office of Science, Basic Energy Sciences, Materials Sciences and Engineering Division. A portion of this research used resources at the High Flux Isotope Reactor and Spallation Neutron Source, which are DOE Office of Science User Facilities operated by the Oak Ridge National Laboratory. The beam time was allocated on WAND$^2$, NOMAD, and CNCS under proposal numbers IPTS-32079, IPTS-31234, and IPTS-34042, respectively. S.D.W. acknowledges support from the DOE Office of Basic Energy Sciences, Division of Materials Sciences and Engineering, under Grant No.~DE-SC0017752. Part of this work utilized facilities at the UC Santa Barbara NSF Quantum Foundry, supported through the Q-AMASE-i program under award DMR-1906325. B.A.F. acknowledges support from the U.S. National Science Foundation, Division of Materials Research, LEAPS-MPS program through Award No. 2418438. We also thank the SNS X-ray Laboratory at ORNL for access to the Laue instrument used in sample coalignment. We are grateful to Matthew S. Powell and Joseph W. Kolis (Clemson University) for providing the La$_2$Sn$_2$O$_7$ single crystal, which was used as a non-magnetic background in the heat capacity measurements reported in this work. 

This manuscript has been authored by UT-Battelle, LLC, under contract DE-AC05-00OR22725 with the U.S. Department of Energy (DOE). The U.S. government retains, and the publisher---by accepting the article for publication---acknowledges that the U.S. government retains a nonexclusive, paid-up, irrevocable, worldwide license to publish or reproduce the published form of this manuscript, or to allow others to do so, for U.S. government purposes. DOE will provide public access to these results of federally sponsored research in accordance with the DOE Public Access Plan (\url{https://www.energy.gov/doe-publicaccess-plan}).

\section{Author Contributions}
Y.L., J.A.M.P., and A.A.A. conceived and designed the project. B.R.O. synthesized the single-crystal samples with assistance from M.J.K. and S.D.W., and performed the bulk characterization shown in Fig.~\ref{fig:Fig3}. Y.L. co-aligned the single-crystal array for neutron experiments. Neutron scattering measurements were carried out by Y.L., J.A.M.P., and A.A.A., with experimental support from S.A.C., M.D.F., A.A.P., and J.L. Y.L. and J.A.M.P. analyzed the single-crystal diffuse-scattering data in Fig.~\ref{fig:Fig2}, while Y.L. analyzed the inelastic neutron-scattering data presented in Fig.~\ref{fig:Fig4} and constructed the conceptual phase diagram in Fig.~\ref{fig:Fig1}. B.A.F. and J.L. performed the structural refinements and pair-distribution-function analyses described in Fig.~\ref{fig:Fig5}. Y.L., J.A.M.P., and A.A.A. wrote the paper with feedback provided by all the other co-authors.

\section{Competing Interests}
The authors declare no competing interests.  

\bibliographystyle{apsrev4-2-titles}
\bibliography{reference}
\clearpage

\onecolumngrid

\begin{center}
\textbf{\large Supplementary Materials for ``Disorder-induced proximate quantum spin ice phase in \texorpdfstring{\pso}{Pr2Sn2O7}''}
\end{center}

\setcounter{figure}{0}
\renewcommand\thetable{S\arabic{table}}

\setcounter{section}{0}
\renewcommand{\thesection}{S\arabic{section}}
\renewcommand{\theHsection}{S\arabic{section}}

\setcounter{subsection}{0}
\renewcommand{\thesubsection}{S\arabic{section}.\arabic{subsection}}
\renewcommand{\theHsubsection}{S\arabic{section}.\arabic{subsection}}

\makeatletter
\renewcommand{\p@subsection}{}
\makeatother

\setcounter{figure}{0}
\renewcommand\thefigure{S\arabic{figure}}

\renewcommand{\theequation}{S\arabic{equation}}
\setcounter{equation}{0}
\renewcommand{\theHequation}{S\arabic{equation}}

\setcounter{table}{0}
\renewcommand{\thetable}{S\arabic{table}}
\renewcommand{\theHtable}{S\arabic{table}}

\section{Supplementary Figures and Tables}\label{Sec:S1_fix}
Unless otherwise stated, uncertainties of data points in all figures correspond to one standard deviation (1$\sigma$) statistical errors. For parameters obtained from $\chi^2$ minimization of the diffuse-scattering fits, uncertainties were determined from the parameter range satisfying $\chi^2 < \chi^2_{\min}\left(1+\frac{1}{N-M}\right)$, where $N$ is the number of data points and $M$ is the number of fitted parameters, taking into account the numerical resolution of the parameter grid. For all other fitted quantities obtained from least-squares fitting procedures, uncertainties were estimated from covariance-matrix analysis near the best-fit solution. Systematic uncertainties are not included.
\begin{figure*}[ht] 
\centering
\includegraphics{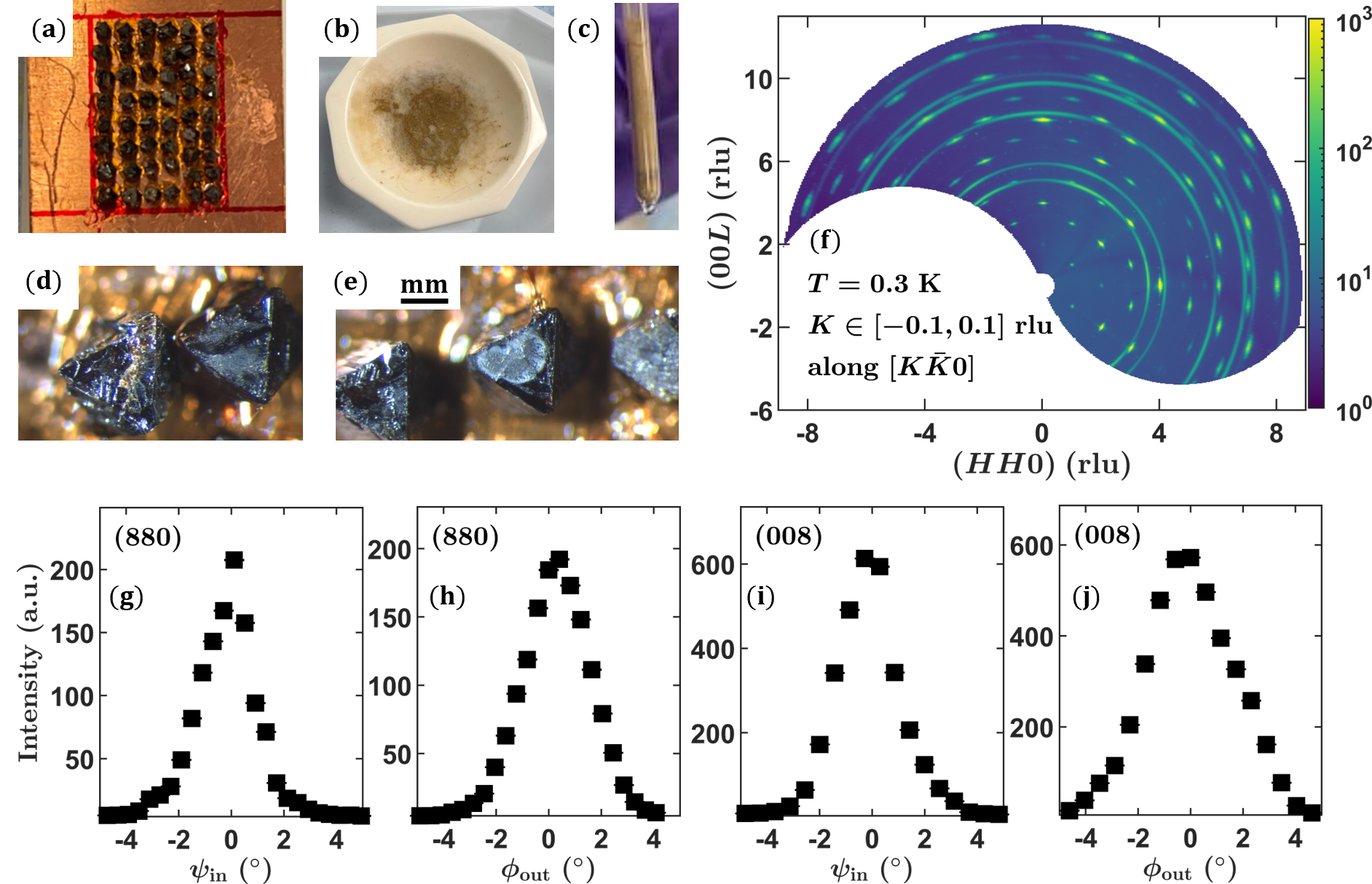}
\caption{\textbf{Sample characterization and mosaicity of the co-aligned single-crystal array.}
(a) Photograph of a portion of the co-aligned single-crystal \pso\ array (total mass $\sim 1.27$~g), with the vertical axis approximately aligned along the $[1\bar{1}0]$ crystallographic direction. 
(b,c) Photographs of the pulverized single-crystal sample used for neutron powder diffraction on the Nanoscale-Ordered Materials Diffractometer (NOMAD), exhibiting a dark red color. 
(d,e) Optical microscopy images of a representative crystal mounted on the copper plate within the co-aligned array shown in (a); the scale bar is indicated in (e). 
(f) Neutron diffraction pattern of the co-aligned array collected at $T = 0.3$~K with incident wavelength $\lambda = 1.5$~\AA\ on the Wide-Angle Neutron Diffractometer (WAND$^2$), indicating a single, slightly broadened crystal grain with an estimated mosaicity of $\sim 5^\circ$ full width at half maximum (FWHM), as quantified in (g--j). 
(g--j) Transverse cuts through the Bragg peaks at $\boldsymbol{Q} = (880)$ and $(008)$ along the in-plane directions ($[00L]$ and $[HH0]$, respectively) and the out-of-plane direction $[K\bar{K}0]$ at $T = 0.3$~K. The horizontal axes are expressed in angular units, $\psi_{\mathrm{in}}$ (in-plane) and $\phi_{\mathrm{out}}$ (out-of-plane), obtained by normalizing to $|\boldsymbol{Q}|$. Data are integrated within $\pm 0.08~\mathrm{\AA}^{-1}$ in the perpendicular $\boldsymbol{Q}$ directions.}
\label{fig:SIFig1}
\end{figure*}
\begin{figure*}[ht] 
\centering
\includegraphics{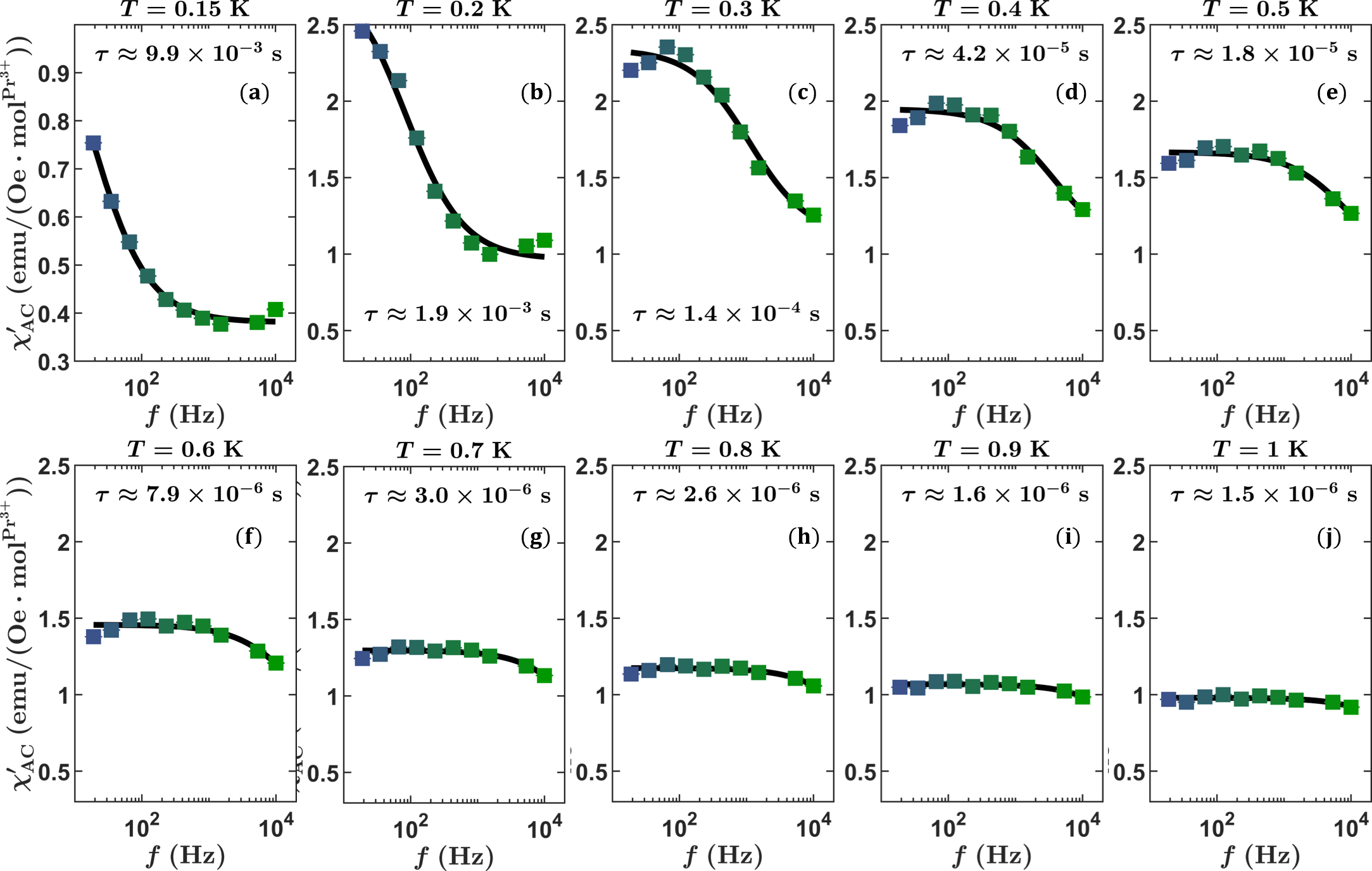}
\caption{\textbf{Frequency-dependent a.c.\ susceptibility.} 
(a--j) Real part $\chi'$ measured as a function of frequency at various temperatures in zero applied d.c.\ field. Solid lines are fits to the damped harmonic oscillator model [Eq.~\ref{dampHarm} in the main text]. The corresponding relaxation times $\tau$ are labeled in each panel. Error bars are omitted for clarity; for $T \geq 0.5$~K, the uncertainties are comparable in magnitude to the extracted $\tau$ values. The extracted timescales should therefore be regarded as order-of-magnitude estimates.}
\label{fig:SIFig2}
\end{figure*}
\begin{figure*}[ht] 
\centering
\includegraphics{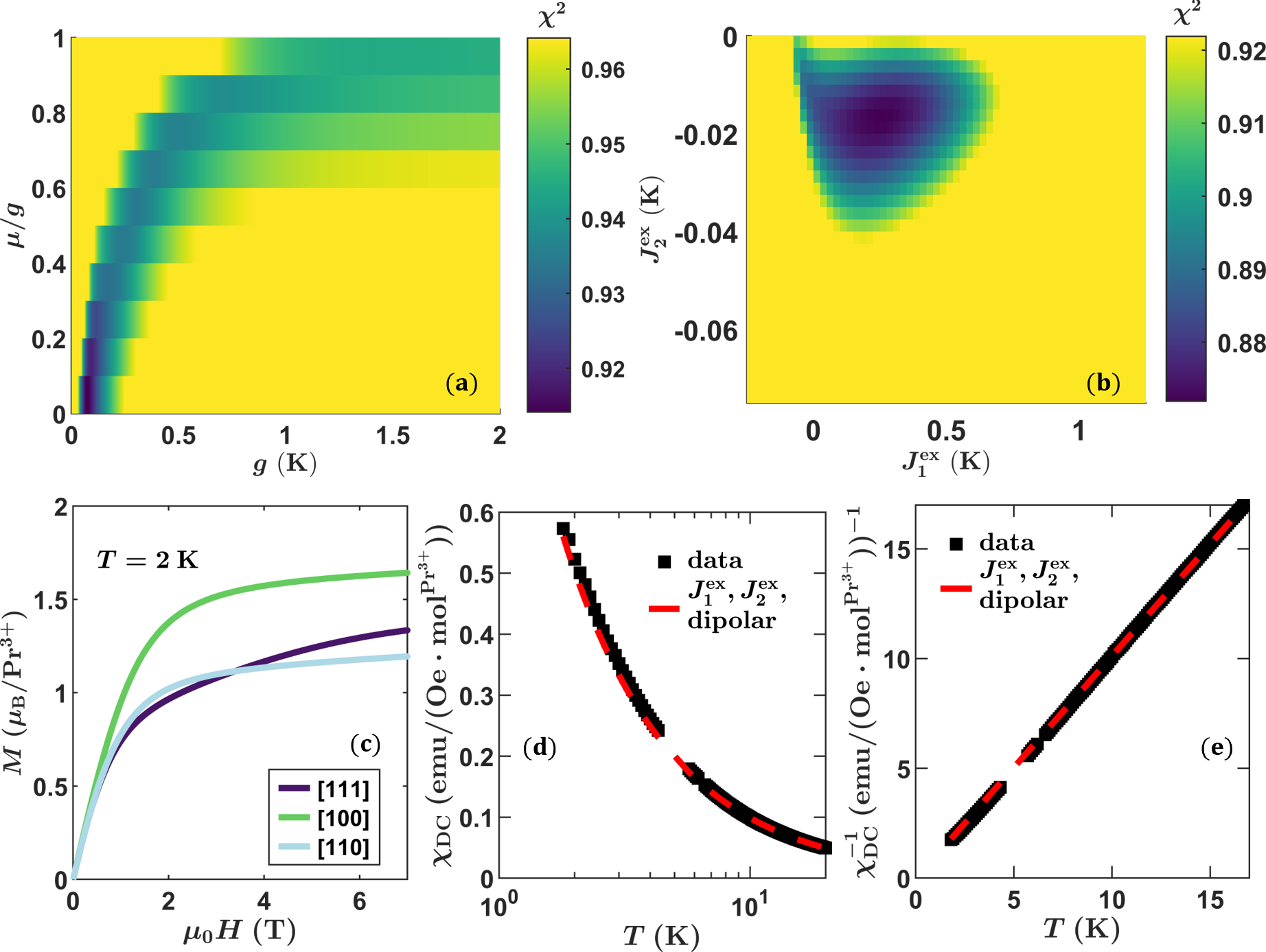}
\caption{\textbf{Model fitting and bulk magnetic characterization.}
(a) Goodness-of-fit map ($\chi^2$) as a function of the ring-exchange parameter $g$ (defined in Eq.~\ref{Hring_const}) and the theoretical tuning parameter $\mu/g$ for the quantum spin-ice model. The quantum limit corresponds to $\mu/g = 0$, where the best-fit value is $g = 0.075(4)$~K$\simeq 0.0065(3)$~meV. 
(b) Goodness-of-fit map ($\chi^2$) as a function of the nearest- and next-nearest-neighbor exchange interactions $J_1^{\mathrm{ex}}$ and $J_2^{\mathrm{ex}}$ [Eq.~\ref{Hc_nnn}] for the dipolar spin-ice model including further-neighbor exchange interactions. The best-fit parameters are $J_1^{\mathrm{ex}} = 0.25(3)$~K$\simeq 0.0215(26)$~meV and $J_2^{\mathrm{ex}} = -0.018(1)$~K$\simeq -0.0016(1)$~meV. 
(c) Magnetization as a function of applied field measured at $T = 2$~K on single-crystal \pso\ samples, with masses of approximately 5.0~mg, 6.3~mg, and 5.6~mg for fields applied along the $[100]$, $[111]$, and $[110]$ directions, respectively. 
(d,e) Temperature dependence of the powder-averaged d.c.\ susceptibility measured on single-crystal \pso, compared with \textsc{SPINTERACT}~\cite{paddison2023spinteract} simulations using the best-fit parameters from panel (b).}
\label{fig:SIFig4}
\end{figure*}
\begin{figure}[ht] 
\centering
\includegraphics{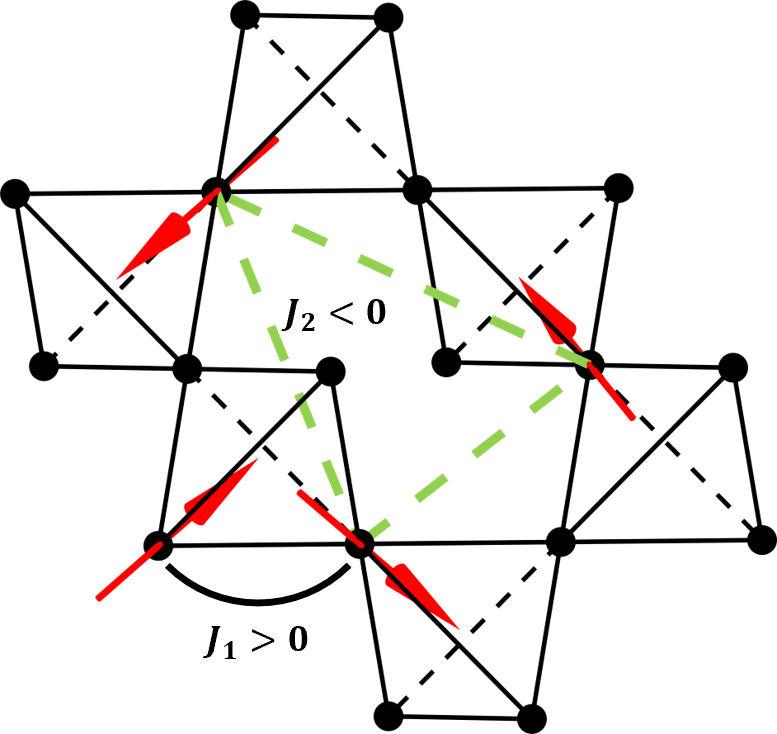}
\caption{\textbf{Sign convention for exchange interactions in the classical spin model.} 
Illustration of the sign convention used for the fitted interaction parameters in the \textsc{SPINTERACT} model~\cite{paddison2023spinteract}. An antiferromagnetic nearest-neighbor interaction $J_1 > 0$ in the local frame favors the canonical ``two-in--two-out'' spin configuration (red arrows) on each tetrahedron. A ferromagnetic next-nearest-neighbor interaction $J_2 < 0$ in the local frame stabilizes a loop-like structure, in which spins on three next-nearest-neighbor sites around a hexagonal plaquette align either clockwise or counterclockwise relative to the plaquette center.}\label{fig:SIpyro}
\end{figure}
\begin{figure*}[t] 
\centering
\includegraphics{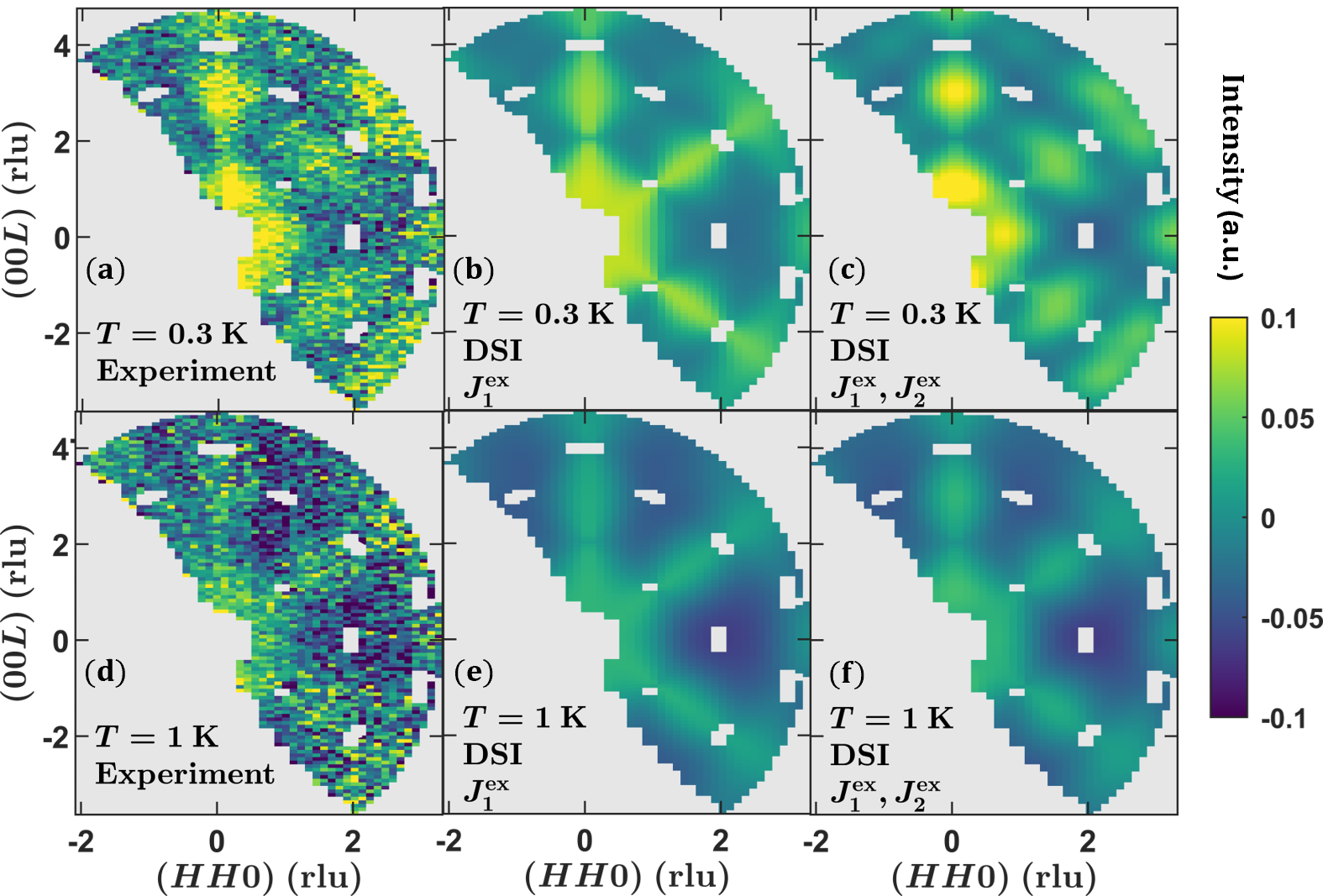}
\caption{\textbf{Supplemental magnetic diffuse scattering and comparison with spin-ice models.}
(a,d) Experimental magnetic diffuse scattering of \pso\ at $T = 0.3$~K and $T = 1$~K, respectively, after subtraction of high-temperature backgrounds ($T = 10$~K for (a) and $T = 20$~K for (d)). Data are integrated along $[K\bar{K}0]$ over $K \in [-0.1,\,0.1]$~reciprocal lattice units (rlu). 
(b,c,e,f) Corresponding simulations using the dipolar spin-ice (DSI) model with long-range dipolar interactions [Eq.~\ref{Hc_nnn}]. Panels (b,e) include nearest-neighbor exchange $J_1^{\mathrm{ex}} = 0.25(3)$~K$\simeq 0.0215(26)$~meV together with dipolar interactions, while panels (c,f) additionally include next-nearest-neighbor exchange $J_2^{\mathrm{ex}} = -0.018(1)$~K$\simeq -0.0016(1)$~meV.}
\label{fig:Fig2_supp}
\end{figure*}

\begin{figure*}[ht] 
\centering
\includegraphics{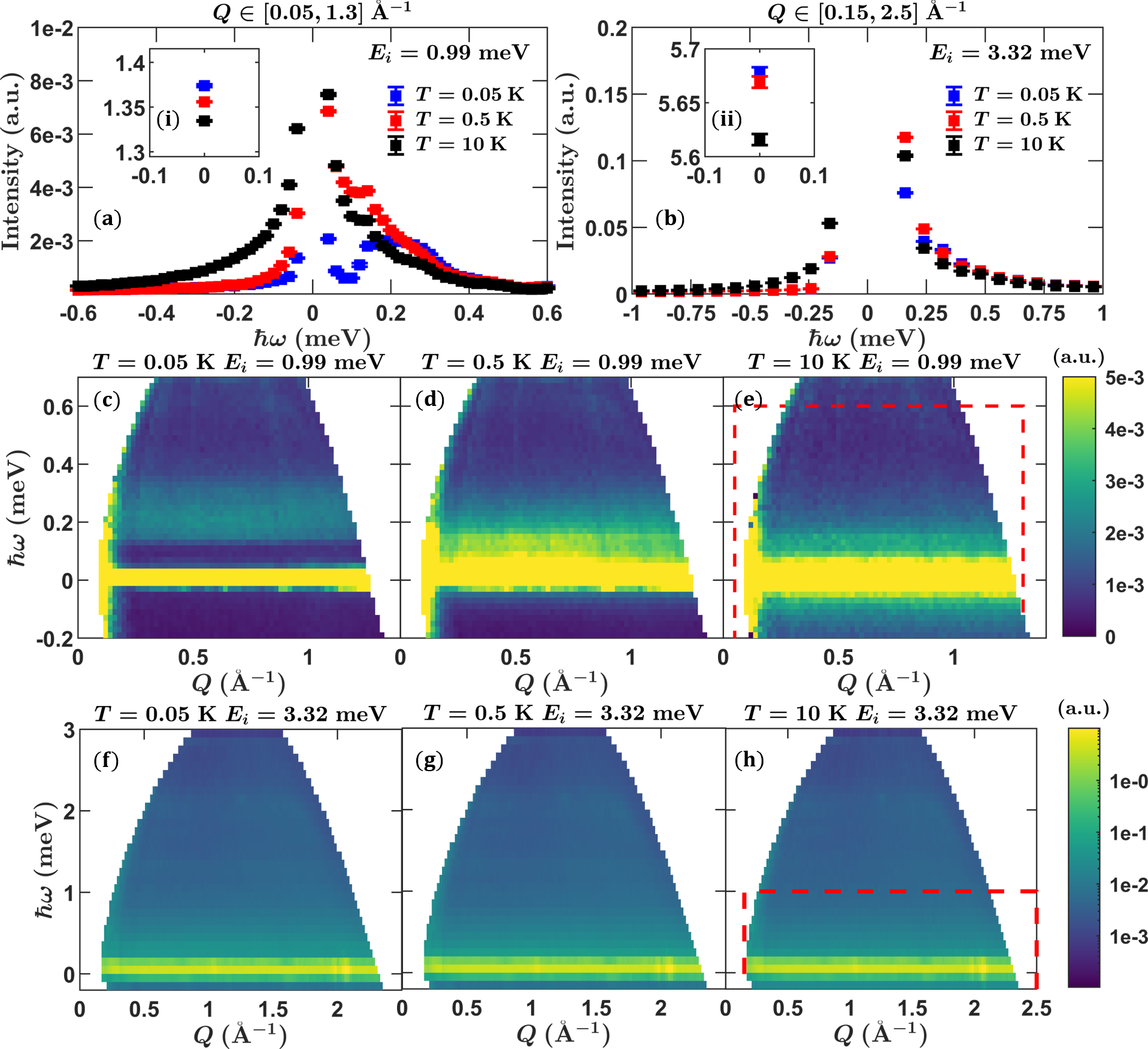}
\caption{\textbf{Powder-averaged neutron scattering spectra at multiple temperatures.} The data shown are raw (no background subtraction).
(a,b) Powder-averaged intensity as a function of energy transfer $\hbar\omega$ at $T = 0.05$, 0.5, and 10~K, measured with incident energies $E_i = 0.99$~meV (a) and $3.32$~meV (b). Data are integrated along $[K\bar{K}0]$ with $K \in [-0.1,\,0.1]$~rlu for $E_i = 0.99$~meV and $K \in [-0.15,\,0.15]$~rlu for $E_i = 3.32$~meV, and averaged over $Q \in [0.05,\,1.3]~\mathrm{\AA}^{-1}$ and $Q \in [0.15,\,2.5]~\mathrm{\AA}^{-1}$, respectively, within the horizontal $(HHL)$ plane. Insets (i,ii) show the corresponding elastic intensities.
(c--h) False-color maps of the powder-averaged intensity as a function of momentum transfer $Q$ and energy transfer $\hbar\omega$. Panels (c--e) correspond to $E_i = 0.99$~meV, while panels (f--h) correspond to $E_i = 3.32$~meV. The red dashed rectangles in panels (e) and (h) indicate the $Q$ ranges used to average the scattering intensity and the energy-transfer windows displayed in panels (a) and (b), respectively.}
\label{fig:SIFig5}
\end{figure*}

\begin{figure*}[t] 
\centering
\includegraphics{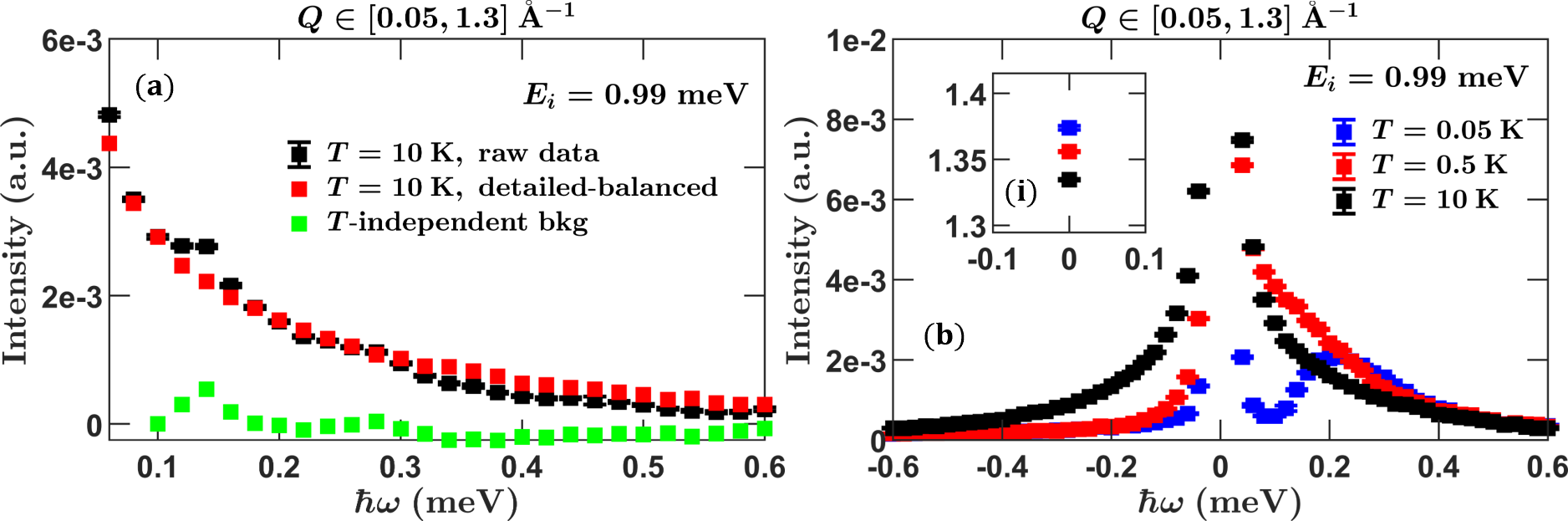}
\caption{\textbf{Background determination and subtraction for inelastic spectra.}
(a) Determination of a shoulder-like background feature at $\hbar\omega \sim 0.1$~meV observed in Fig.~\ref{fig:SIFig5}(a), measured with $E_i = 0.99$~meV. The black curve shows the raw inelastic scattering intensity on the energy-loss side ($\hbar\omega > 0$) at $T = 10$~K. The red curve represents the corresponding intensity calculated from the energy-gain side using the detailed balance relation $I(|\hbar\omega|) = I(-|\hbar\omega|)\exp(|\hbar\omega|/k_{\mathrm{B}} T)$. The green curve shows the difference between the measured (black) and calculated (red) curves, providing an estimate of the temperature-independent background. 
(b,i) Background-subtracted version of the powder-averaged spectra shown in Fig.~\ref{fig:SIFig5}(a,i). The shoulder-like background determined in panel (a) is subtracted from all three temperatures.}
\label{fig:SIFig6}
\end{figure*}

\begin{figure*}[ht] 
\centering
\includegraphics{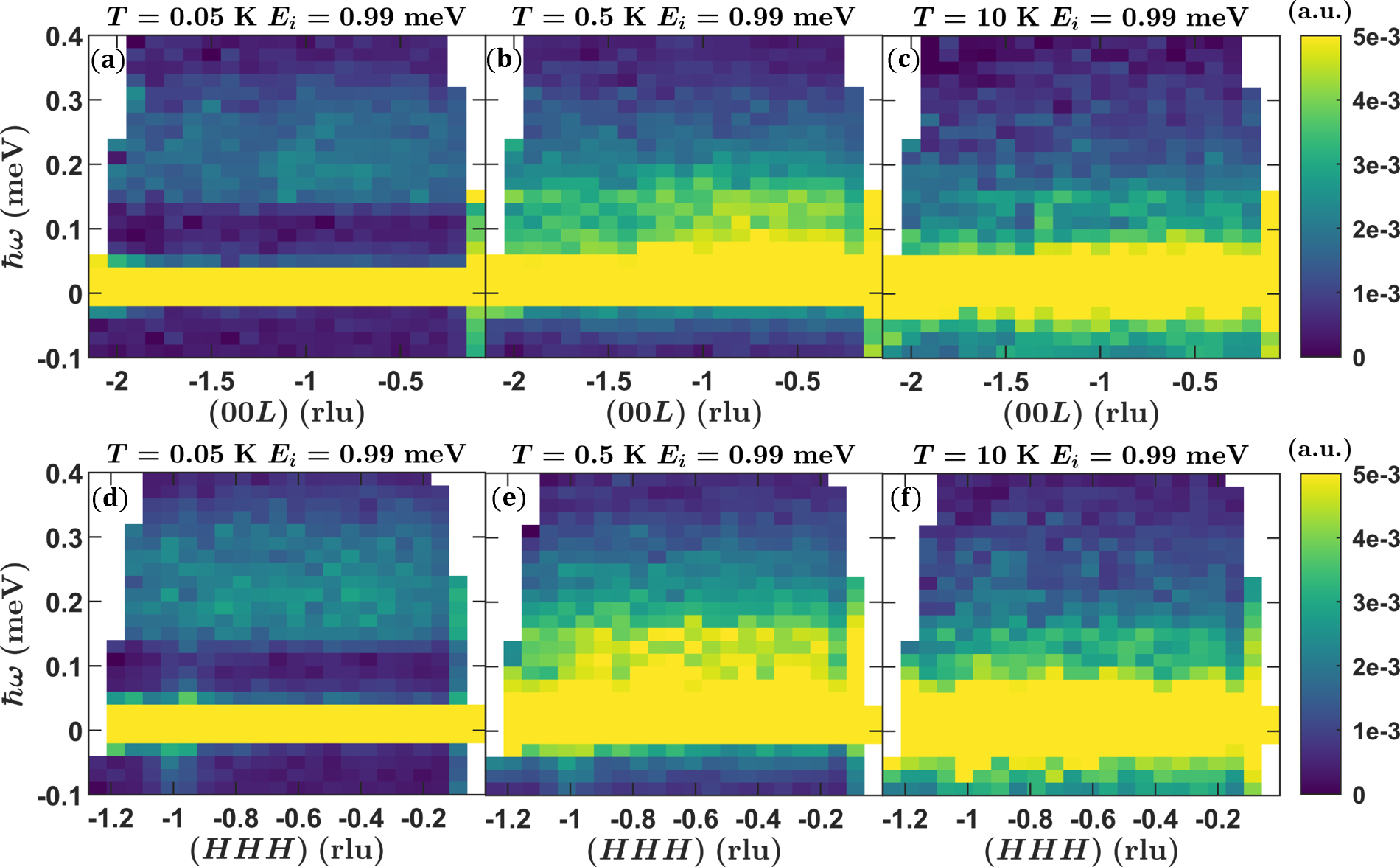}
\caption{\textbf{Momentum-resolved inelastic scattering along high-symmetry directions.}
(a--c) False-color maps of the scattering intensity as a function of momentum transfer along the $[00L]$ direction and energy transfer $\hbar\omega$, measured at $T = 0.05$~K, $0.5$~K, and $10$~K, respectively. 
(d--f) Corresponding maps along the $[HHH]$ direction as a function of $\hbar\omega$ at the same temperatures. 
All data were collected with incident energy $E_i = 0.99$~meV and integrated over $\pm 0.15$~rlu in directions perpendicular to the respective cut directions.}
\label{fig:SIFig7}
\end{figure*}

\begin{figure*}[ht] 
\centering
\includegraphics{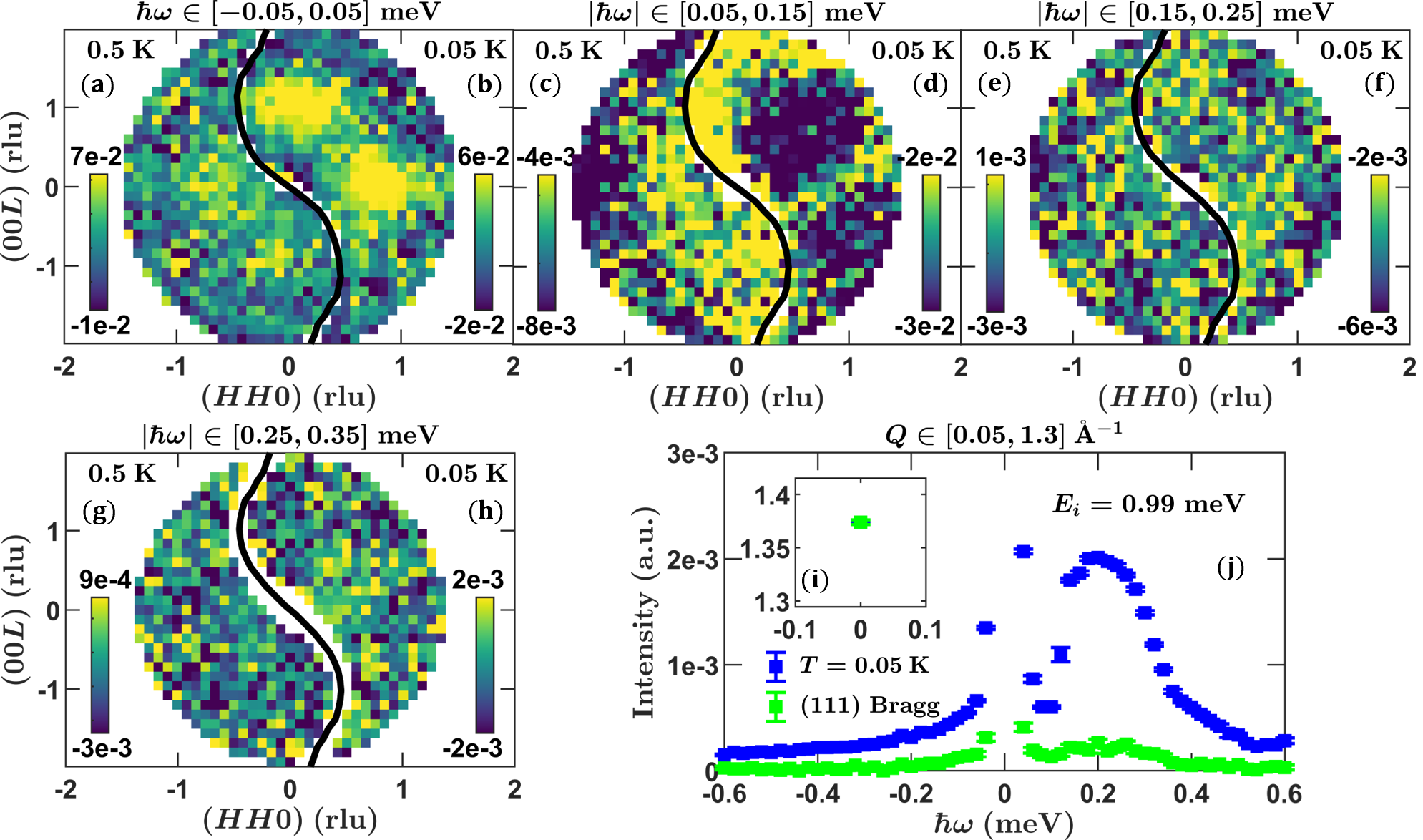}
\caption{\textbf{Low-energy $\boldsymbol{Q}$-dependent scattering and resolution characterization.}
(a--h) Constant-energy $\boldsymbol{Q}$ maps (arb.\ units) measured at $T = 0.5$~K (a,c,e,g) and $T = 0.05$~K (b,d,f,h) using incident energy $E_i = 0.99$~meV (elastic energy resolution FWHM $\approx 0.02$~meV), following subtraction of the $T = 10$~K background. Data are integrated along $[K\bar{K}0]$ with $K \in [-0.1,\,0.1]$~rlu, and over energy ranges $\hbar\omega \in [-0.05,\,0.05]$~meV (a,b), $[0.05,\,0.15]$~meV (c,d), $[0.15,\,0.25]$~meV (e,f), and $[0.25,\,0.35]$~meV (g,h). No symmetrization is applied. Color scales are centered on the mean intensity of each panel. Energy-gain and energy-loss contributions are combined in panels (c--h). 
(j) Powder-averaged intensity as a function of energy transfer $\hbar\omega$ at $T = 0.05$~K for scattering near the horizontal $(HHL)$ plane ($K \leq 0.1$~rlu along $[K\bar{K}0]$), measured with $E_i = 0.99$~meV. The green solid curve shows the intensity integrated within $\pm 0.1$~rlu around the (111) Bragg peak, scaled to match the elastic intensity in (i), and serves as an in-situ measure of the instrumental energy resolution.}
\label{fig:SIFig8}
\end{figure*}

\begin{figure*}[h!] 
\centering
\includegraphics{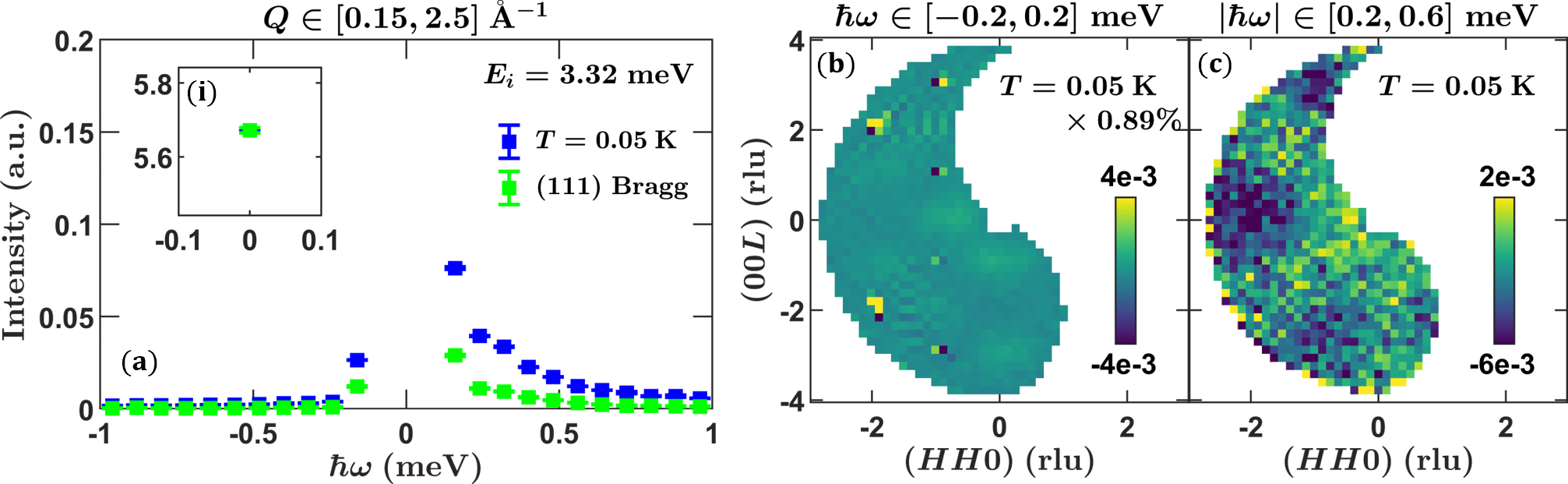}
\caption{\textbf{Resolution analysis and assessment of elastic leakage in inelastic constant energy slices.}
(a) Powder-averaged scattering intensity as a function of energy transfer $\hbar\omega$ at $T = 0.05$~K, measured near the horizontal $(HHL)$ plane ($K \leq 0.15$~rlu along $[K\bar{K}0]$) with incident energy $E_i = 3.32$~meV. The green solid curve shows the intensity integrated within $\pm 0.15$~rlu around the (111) Bragg peak and scaled to match the elastic peak intensity in inset (i), providing an in-situ measure of the instrumental energy resolution. Approximately $99.49\%$ of the integrated intensity of the (111) Bragg peak lies within the energy window $\hbar\omega \in [-0.2,\,0.2]$~meV. 
(b,c) Constant-energy $\boldsymbol{Q}$ maps (arb.\ units) at $T = 0.05$~K, measured with $E_i = 3.32$~meV after subtraction of the $T = 10$~K background. Data are integrated along $[K\bar{K}0]$ with $K \in [-0.15,\,0.15]$~rlu and over energy ranges $\hbar\omega \in [-0.2,\,0.2]$~meV (b) and $[0.2,\,0.6]$~meV (c). No symmetrization is applied. Color scales in (b) and (c) are centered on the mean intensity and span the same symmetric range about the mean. In (c), energy-gain and energy-loss contributions are combined. The intensity in (b) is scaled by $1 - 99.49\% = 0.51\%$ to estimate the expected elastic leakage into the $[0.2,\,0.6]$~meV window due to finite resolution, demonstrating that the ``star-fish-like'' $\boldsymbol{Q}$ dependence observed in (c) cannot be attributed to elastic leakage.}
\label{fig:SIFig9}
\end{figure*}

\begin{figure*}[h!] 
\centering
\includegraphics{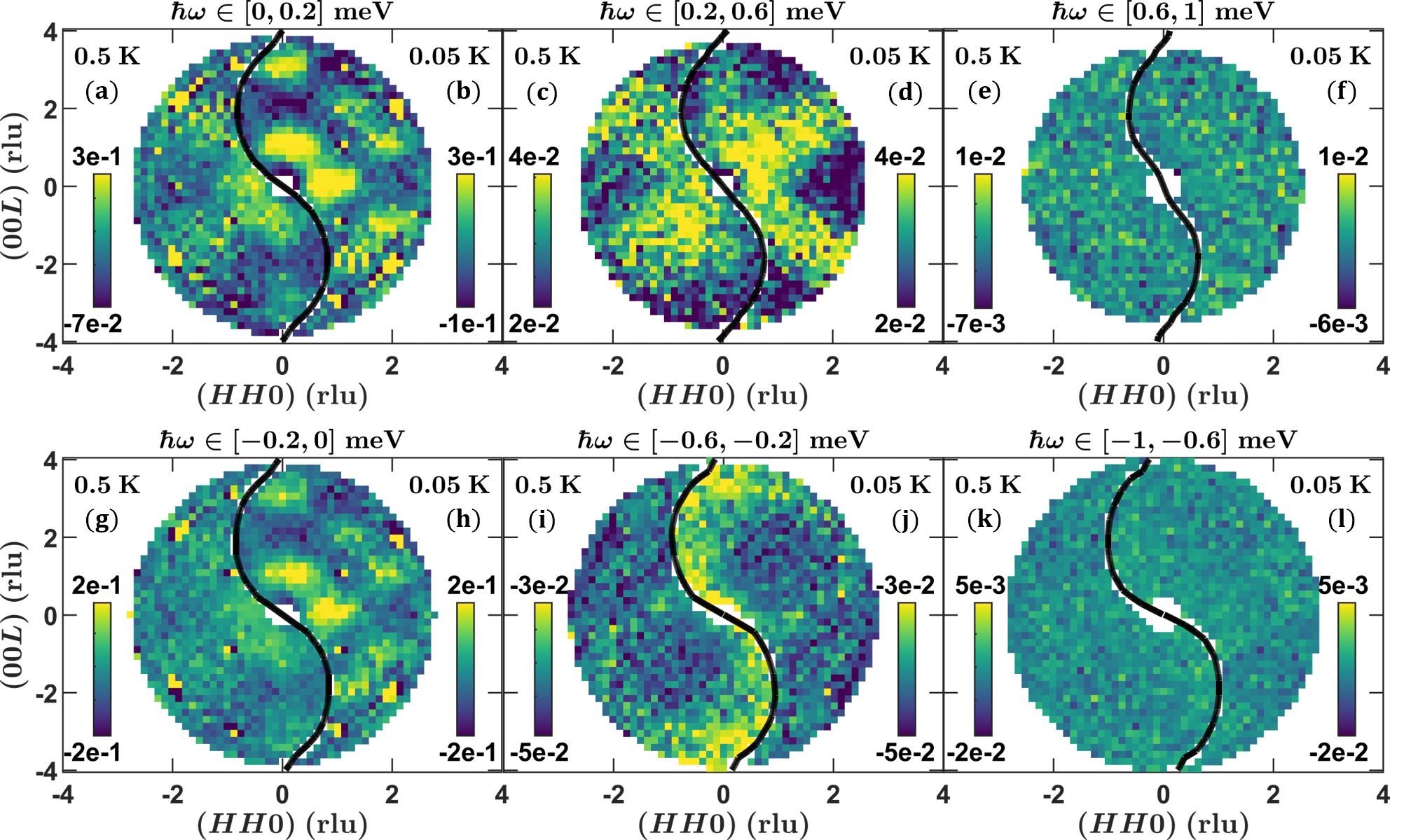}
\caption{\textbf{Extended constant-energy $\boldsymbol{Q}$ maps at multiple temperatures.}
Constant-energy $\boldsymbol{Q}$ maps (arb.\ units) measured at $T = 0.5$~K (a,c,e,g,i,k) and $T = 0.05$~K (b,d,f,h,j,l), using incident energy $E_i = 3.32$~meV (elastic energy resolution FWHM $\approx 0.1$~meV) after subtraction of the $T = 10$~K background. Data are integrated along $[K\bar{K}0]$ with $K \in [-0.15,\,0.15]$~rlu and over the energy ranges indicated in each panel. No symmetrization is applied. Color scales are centered on the mean intensity of each panel.}
\label{fig:SIFig10}
\end{figure*}

\begin{table*}[ht]
\centering
\caption{\textbf{Refined atomic positions and anisotropic displacement parameters.}
Atomic positions and anisotropic displacement parameters for pulverized single-crystal \pso, determined from neutron powder diffraction data collected on NOMAD and neutron PDF analysis under the assumption of full stoichiometry.}\label{tab:refine}

\subfloat[Rietveld refinement from data collected at $T=100$ K.]{
\begin{tabular}{lccccccccc}
\hline\hline
\multicolumn{5}{l}{Lattice parameter (\AA): 10.59999(8)} & \multicolumn{5}{r}{Space group: \textit{Fd$\bar{3}$m}} \\
\hline
Atom (Wyckoff) & $x$ & $y$ & $z$ & $u_{11}$ & $u_{22}$ & $u_{33}$ & $u_{12}$ & $u_{13}$ & $u_{23}$ \\
\hline
Sn (16c) & 0 & 0 & 0 & 0.00072(7) & 0.00072(7) & 0.00072(7) & 0.00005(7) & 0.00005(7) & 0.00005(7) \\
Pr (16d) & 0.5 & 0.5 & 0.5 & 0.00259(10) & 0.00259(10) & 0.00259(10) & -0.00046(11) & -0.00046(11) & -0.00046(11) \\
O (8b)   & 0.375 & 0.375 & 0.375 & 0.00258(10) & 0.00258(10) & 0.00258(10) & 0 & 0 & 0 \\
O (48f)  & 0.331480(18) & 0.125 & 0.125 & 0.00264(10) & 0.00337(8) & 0.00337(8) & 0 & 0 & 0.00141(9) \\
\hline\hline
\end{tabular}
}\label{NOM_100K}

\subfloat[Rietveld refinement from data collected at $T=300$ K.]{
\begin{tabular}{lccccccccc}
\hline\hline
\multicolumn{5}{l}{Lattice parameter (\AA): 10.60596(9)} & \multicolumn{5}{r}{Space group: \textit{Fd$\bar{3}$m}} \\
\hline
Atom (Wyckoff) & $x$ & $y$ & $z$ & $u_{11}$ & $u_{22}$ & $u_{33}$ & $u_{12}$ & $u_{13}$ & $u_{23}$ \\
\hline
Sn (16c) & 0 & 0 & 0 & 0.00180(9) & 0.00180(9) & 0.00180(9) & 0.00028(8) & 0.00028(8) & 0.00028(8) \\
Pr (16d) & 0.5 & 0.5 & 0.5 & 0.00507(13) & 0.00507(13) & 0.00507(13) & -0.00102(12) & -0.00102(12) & -0.00102(12) \\
O (8b)   & 0.375 & 0.375 & 0.375 & 0.00400(13) & 0.00400(13) & 0.00400(13) & 0 & 0 & 0 \\
O (48f)  & 0.331219(19) & 0.125 & 0.125 & 0.00452(12) & 0.00537(10) & 0.00537(10) & 0 & 0 & 0.00252(10) \\
\hline\hline
\end{tabular}
}\label{NOM_300K}

\subfloat[PDF refinement from data collected at $T=100$ K.]{
\begin{tabular}{lccccccccc}
\hline\hline
\multicolumn{5}{l}{Lattice parameter (\AA): 10.5952(3)} & \multicolumn{5}{r}{Space group: \textit{Fd$\bar{3}$m}} \\
\hline
Atom (Wyckoff) & $x$ & $y$ & $z$ & $u_{11}$ & $u_{22}$ & $u_{33}$ & $u_{12}$ & $u_{13}$ & $u_{23}$ \\
\hline
Sn (16c) & 0 & 0 & 0 & 0.00167(4) & 0.00167(4) & 0.00167(4) & -0.00022(7) & -0.00022(7) & -0.00022(7) \\
Pr (16d) & 0.5 & 0.5 & 0.5 & 0.00285(7) & 0.00285(7) & 0.00285(7) &
-0.00027(12) &
-0.00027(12) &
-0.00027(12) \\
O (8b)   & 0.375 & 0.375 & 0.375 & 0.00353(10) & 0.00353(10) & 0.00353(10) & 0 & 0 & 0 \\
O (48f)  & 0.331348(52) & 0.125 & 0.125 & 0.00465(12) & 0.00396(8) & 0.00396(8) & 0 & 0 & 0.00144(9) \\
\hline\hline
\end{tabular}
}\label{PDF_100K}

\subfloat[Selected bond lengths and bond angles obtained from the PDF refinement assuming the ideal pyrochlore structural model at $T=100$ K.]{

\begin{tabular}{lc|lc}
\hline\hline
Bond & Length (\AA) & Bond angle & Angle ($^\circ$) \\
\hline
Pr--O1 (48f) & 2.5887 & Pr--O1--Pr & 92.695 \\
Pr--O2 (8b)  & 2.2939 & Pr--O2--Pr & 109.471 \\
Sn--O         & 2.0618 & Sn--O--Sn  & 130.58 \\
Pr--Pr, Sn--Sn, Pr--Sn & 3.7460 & Pr--O--Sn & 106.77 \\
\hline\hline
\end{tabular}
}\label{PDF_bond}
\end{table*}
\clearpage

\section{Supplementary bulk characterization of \texorpdfstring{\pso}{Pr2Sn2O7}}\label{APP:sample}
Magnetization versus field measurements at $T = 2$~K on single-crystal \pso\ samples with masses of approximately 5~mg are shown in Fig.~\ref{fig:SIFig4}c. The induced magnetic moments $M$ at the maximum applied field $\mu_0H = 7$~T are measured to be $1.64~\mu_\mathrm{B}/\mathrm{Pr^{3+}}$, $1.33~\mu_\mathrm{B}/\mathrm{Pr^{3+}}$, and $1.19~\mu_\mathrm{B}/\mathrm{Pr^{3+}}$ for fields applied along the $[100]$, $[111]$, and $[110]$ directions, respectively. For comparison, the expected saturated moments in a pyrochlore lattice with Ising anisotropy along the local $\langle 111 \rangle$ directions and effective ferromagnetic (FM) coupling between nearest neighbors are $g_J J /\sqrt{3} $ for the $[100]$ direction, $g_J J /2 $ for $[111]$, and $g_J J /\sqrt{6} $ for $[110]$~\cite{Fukazawa2002}. Both the observed ordering of the induced moments (from largest to smallest) and their relative ratios are consistent with the predictions for a magnetic pyrochlore with nearest neighbor FM coupling and easy-axis anisotropy. Quantitative deviations may arise from slight misalignments during sample mounting, mass uncertainty due to residual impurity phases, or the fact that full saturation may not have been achieved at $\mu_0H = 7$~T. The temperature dependence of the d.c. magnetic susceptibility was measured with the magnetic field applied along the $[100]$, $[111]$, and $[110]$ crystallographic directions using the same single-crystal samples. To obtain a representative bulk behavior, we compute a powder-averaged d.c. susceptibility for a cubic system using the expression $\chi_{\mathrm{powder}}^{\mathrm{DC}} = (3\chi_{100}^{\mathrm{DC}} + 6\chi_{110}^{\mathrm{DC}} + 4\chi_{111}^{\mathrm{DC}})/13$ \cite{bonville2016magnetic}, and plot the results in Fig.~\ref{fig:SIFig4}d,e. This powder-averaged d.c. susceptibility is also used to normalize the real part of the a.c. susceptibility $\chi'$ presented in Fig.~3a of the main text. A Curie–Weiss fit to the high-temperature regime yields a Curie–Weiss temperature of $\theta_{\mathrm{CW}} = -0.23(2)$~K and an effective moment of $\mu_{\mathrm{eff}} = 2.78(1)~\mu_\mathrm{B}/\mathrm{Pr^{3+}}$, in good agreement with the powder measurements reported in Ref.~\cite{Ortiz2024Pr2Sn2O7}.

\section{Modeling of magnetic diffuse-scattering data}\label{APP:diffuse}
To model the magnetic diffuse scattering observed in single-crystal \pso, we use the \textsc{SPINTERACT} software suite~\cite{paddison2023spinteract}. For quantum and classical models, we write the Hamiltonian in terms of unit Pauli matrices or unit-length Ising variables, respectively, corresponding to an effective spin-$1/2$ on each Pr$^{3+}$ site with an effective $g$-factor $g_{\parallel} = 5.17$ along the local $\langle111\rangle$ direction~\cite{Princep2013Pr2Sn2O7_CrystalField}. The local coordinate frames $(\hat{\mathbf{a}}_i,\hat{\mathbf{b}}_i,\hat{\mathbf{e}}_i)$ are defined as in Ref.~\cite{Ross2011}:
\begin{align}
\begin{split}
\hat{\mathbf{e}}_0 &= \frac{1}{\sqrt{3}}(1,1,1),~~~~ ~~~\hat{\mathbf{a}}_0 = \frac{1}{\sqrt{6}}(-2,1,1),\\
\hat{\mathbf{e}}_1 &= \frac{1}{\sqrt{3}}(1,-1,-1),\quad \hat{\mathbf{a}}_1 = \frac{1}{\sqrt{6}}(-2,-1,-1),\\
\hat{\mathbf{e}}_2 &= \frac{1}{\sqrt{3}}(-1,1,-1),\quad \hat{\mathbf{a}}_2 = \frac{1}{\sqrt{6}}(2,1,-1),\\
\hat{\mathbf{e}}_3 &= \frac{1}{\sqrt{3}}(-1,-1,1),\quad \hat{\mathbf{a}}_3 = \frac{1}{\sqrt{6}}(2,-1,1),
\end{split}
\label{loc_cord}
\end{align}
corresponding to the four Pr$^{3+}$ ions located at $\hat{\mathbf{r}}_0=(0.5, 0.5, 0.5)$, $\hat{\mathbf{r}}_1=(0.5, 0.75, 0.75)$, $\hat{\mathbf{r}}_2=(0.75, 0.5, 0.75)$, and $\hat{\mathbf{r}}_3=(0.75, 0.75, 0.5)$ in the face-centered cubic unit cell, with $\hat{\mathbf{b}}_i=\hat{\mathbf{e}}_i\times\hat{\mathbf{a}}_i$. A magnetic background following the form factor $|F(Q)|^2$ for Pr$^{3+}$ is subtracted to simulate the experimental high-temperature background removal. The \textsc{SPINTERACT} program minimizes the goodness-of-fit metric $\chi^2$, defined as
\begin{equation}
\begin{split}
\chi^2 &= \sum_{d} W_d \sum_{i \in d} \left( \frac{I_i^{\text{data}} - s_d I_i^{\text{calc}}}{\sigma_i} \right)^2, \\
s_d &= \frac{\sum_i I_i^{\text{data}} I_i^{\text{calc}} / \sigma_i^2}{\sum_i (I_i^{\text{calc}})^2 / \sigma_i^2},
\end{split}\label{eqn:chi2def}
\end{equation}
where $d$ indexes a particular dataset with weight $W_d$, $I_i^{\text{data}}$ and $I_i^{\text{calc}}$ denote the measured and calculated intensities for the $i$-th data point, respectively, and $\sigma_i$ is the corresponding experimental uncertainty. The scale factor $s_d$ for each dataset is determined via the linear least-squares expression given above. Throughout this work, the $\chi^2$ values reported in Fig.~\ref{fig:SIFig4}(a,b) correspond to the reduced goodness-of-fit metric, obtained by dividing the $\chi^2$ quantity defined above in Eq.~\ref{eqn:chi2def} by the number of residual degrees of freedom $(N-M)$, consistent with the uncertainty criterion described at the beginning of Sec.~\ref{Sec:S1_fix}.

For the quantum model, we adopt the theoretical results of Ref.~\cite{Benton2012}, where the goodness-of-fit $\chi^2$ is plotted against the ring-exchange parameter $g$ [Fig.~\ref{fig:SIFig4}a] in the effective Hamiltonian:
\begin{equation}
\mathcal{H}_{\text{ring}} = -\sum_{\hexagon}
g\left( \sigma^{+}_i \sigma^{-}_j \sigma^{+}_k \sigma^{-}_l \sigma^{+}_m \sigma^{-}_n + \text{H.c.} \right).
\label{Hring_const}
\end{equation}
Here, $\mu$ is a theoretical tuning parameter, with $\mu/g = 0$ corresponding to the quantum spin ice (QSI) limit and $\mu/g = 1$ to the classical limit. This model assumes perturbative interactions with nearest-neighbor exchange $J_1 \rightarrow \infty$. Best-fit analysis yields $g = 0.075(4)$~K$\simeq0.0065(3)$~meV in the QSI limit ($\mu/g = 0$, see Fig.~2b,f in the main text). In the classical limit ($\mu/g = 1$), the scattering profile is independent of $g$, and the result corresponds to the $J_1$-only classical nearest-neighbor spin-ice (NNSI) scenario shown in Fig.~2c,g in the main text. The decreasing trend of $\chi^2$ with increasing $g$ in this regime is attributed to numerical artifacts.

For the classical dipolar spin-ice (DSI) model, we employ Onsager reaction-field theory, incorporating long-range dipolar interactions $D_{\mathrm{dip}}=\mu_0\mu_{\mathrm{B}}^2/(4\pi k_{\mathrm{B}}|\mathbf{r}_{nn}|^3) \approx 0.012$~K$\simeq 0.0010$~meV, following the convention of Ref.~\cite{paddison2023spinteract}. The total interaction $J_n$ between $n$-th neighbor $\sigma_i^z$ terms in Eq.~1 in the main text thus includes both dipolar and exchange contributions,  
\begin{equation}
J_n = D_n + J_n^{\mathrm{ex}}.
\end{equation}
Further-neighbor interactions, in particular $J_2$ and $J_3$, have also been invoked to explain the magnetic correlations observed below the spin-freezing transition in \dto~\cite{Yavorskii2008EmergentClusters,Samarakoon2022Glassiness}. The Hamiltonian, including exchange interactions up to next-nearest neighbors ($J_2$) and neglecting disorder terms, is given by
\begin{equation}
\begin{split}
\mathcal{H} &= \sum_{\langle ij \rangle_1} (J_1^{\mathrm{ex}} + D_1)\, \sigma_i^z \sigma_j^z
+ \sum_{\langle ij \rangle_2} (J_2^{\mathrm{ex}} + D_2)\, \sigma_i^z \sigma_j^z \\
&\quad + \text{further-neighbor dipolar terms},
\label{Hc_nnn}
\end{split}
\end{equation}
where $D_1=\frac{5}{3}(g_{\parallel}S)^2D_{\mathrm{dip}} = 0.13$~K $\simeq 0.011$~meV ($S=1/2$)~\cite{DenHertog2000} is the effective nearest-neighbor dipolar coupling. The next-nearest-neighbor dipolar coupling follows from the corresponding local-axis geometry as $D_2=-D_1/(5\sqrt{3}) = -0.015$~K $\simeq -0.0013$~meV. All parameters are expressed in the local coordinate frames defined in Eq.~\ref{loc_cord} and the long-ranged nature of the dipolar interaction was implemented using Ewald summation. The magnetic diffuse scattering pattern alone does not effectively constrain the magnitude of $J_1^{\mathrm{ex}}$, so we complement the fitting with powder-averaged d.c. susceptibility data ($W_d=1$ for all data sets). A $\chi^2$ map of the fits to the magnetic diffuse scattering and d.c. susceptibility data as a function of $J_1^{\mathrm{ex}}$ and $J_2^{\mathrm{ex}}$ is shown in Fig.~\ref{fig:SIFig4}b, with the optimal parameters found to be $J_1^{\mathrm{ex}} = 0.25(3)$~K$\simeq 0.0215(26)$~meV and $J_2^{\mathrm{ex}} = -0.018(1)$~K$\simeq -0.0016(1)$~meV. We estimate an effective nearest-neighbor interaction of $J_1 = J_1^{\mathrm{ex}} + D_1 \approx 0.033$~meV, which is used in the Discussion section of the main text. The corresponding fits to the susceptibility data are shown in Fig.~\ref{fig:SIFig4}d,e. 

The sign convention for the fitted interaction parameters is illustrated in Fig.~\ref{fig:SIpyro}. An antiferromagnetic nearest-neighbor interaction $J_1 = J_1^{\mathrm{ex}} + D_1 > 0$ between pseudo-spins in the local frame favors the canonical ``two-in, two-out'' spin configuration on each tetrahedron. In contrast, a ferromagnetic next-nearest-neighbor interaction $J_2 = J_2^{\mathrm{ex}} + D_2 < 0$ in the local frame favors a ``loop'' structure, where spins on three next-nearest-neighbor sites around a hexagonal plaquette align either all clockwise or all counterclockwise relative to the plaquette center. 

We note that a model including only $J_1^{\mathrm{ex}}$ and dipolar interactions fails to reproduce the observed $\boldsymbol{Q}$ dependence at $T = 0.3$~K (see Fig.~\ref{fig:Fig2_supp}a,b). Incorporating a finite next-nearest-neighbor exchange $J_2^{\mathrm{ex}}$ is necessary to capture the $\boldsymbol{Q}$ dependence [Fig.~\ref{fig:Fig2_supp}c].

\section{Supplementary neutron spectroscopy analysis}\label{APP:CNCS}

In this section, we present supplementary neutron spectroscopy data and associated analysis for single-crystal \pso. These include additional figures as well as technical aspects of data processing, including background subtraction, energy resolution considerations, and the treatment of energy-gain and energy-loss channels.

\subsection{Supplementary neutron spectral data}

Figure~\ref{fig:SIFig5} shows the powder-averaged scattering intensity as a function of energy transfer $\hbar\omega$. For measurements with $E_i = 0.99$~meV, a gapped inelastic continuum centered at $\hbar\omega \approx 0.2$~meV is observed at $T = 0.05$~K, while a broad low-energy continuum extending up to $\hbar\omega \approx 0.6$~meV is observed at $T = 0.5$~K [Fig.~\ref{fig:SIFig5}a,c,d]. In contrast, the $E_i = 3.32$~meV data show no additional inelastic features above $\sim 0.6$~meV [Fig.~\ref{fig:SIFig5}f,g]. Figure~\ref{fig:SIFig7} presents false-color maps of the scattering intensity as functions of momentum transfer along the $[00L]$ and $[HHH]$ directions and energy transfer, measured with $E_i = 0.99$~meV. These maps exhibit spectral features consistent with those observed in the powder-averaged data [Fig.~\ref{fig:SIFig5}]. 

\subsection{Removal of background feature}\label{APP:sub}

In the raw data shown in Fig.~\ref{fig:SIFig5}a, we identify a temperature-independent, shoulder-like background feature centered near $\hbar\omega \sim 0.1$~meV. This feature is absent on the energy-gain side ($\hbar\omega < 0$) at $T = 10$~K. To remove this contribution, we assume that the background is temperature independent and that the energy-gain side contains negligible background. This assumption is supported by the absence of significant intensity for $Q > 0.2$~\AA$^{-1}$ at $\hbar\omega < 0$ and $T = 0.05$~K (Fig.~\ref{fig:SIFig5}c), where true inelastic scattering is expected to be strongly suppressed.

As illustrated in Fig.~\ref{fig:SIFig6}a, we use the energy-gain-side data at $T = 10$~K to calculate the corresponding energy-loss spectrum ($\hbar\omega > 0$) via the detailed balance relation $I(|\hbar\omega|) = I(-|\hbar\omega|)\exp(|\hbar\omega|/k_{\mathrm{B}} T)$ (red curve). The asymmetry in energy resolution between the energy-gain and energy-loss sides is neglected, as the relevant energy range lies close to the elastic line. The background contribution (green curve) is then obtained by subtracting the calculated energy-loss spectrum (red) from the measured data (black). This estimated background is subtracted from the spectra at all three temperatures, yielding the corrected data shown in Fig.~\ref{fig:SIFig6}b, which are used for the quantitative analysis in Fig.~3a,b,c of the main text.

% note on 07/02/2026: in the next round of revision, consider adding the description regarding reviewer #1, comment 7, about detailed-balance date in Fig. S7(a) slightly higher than raw data in positive energy transfer >0.3 meV.

\subsection{Energy resolution and elastic leakage analysis}\label{APP:CNCS_ConstE}

In Fig.~\ref{fig:SIFig8}a--h, we present constant-energy $\boldsymbol{Q}$ maps analogous to those shown in Fig.~4d--j, but measured with an incident energy of $E_i = 0.99$~meV. This configuration provides superior energy resolution, albeit at the expense of reduced neutron flux and more limited $\boldsymbol{Q}$ coverage. At $T = 0.05$~K, signatures of incipient (100) correlations are observed as diffuse scattering centered near the $(001)$ and $(110)$ positions (Fig.~\ref{fig:SIFig8}b). However, owing to limited statistics and restricted $\boldsymbol{Q}$ range, the momentum dependence of the low-energy (quasielastic and inelastic) response at $T = 0.5$~K, as well as the gapped inelastic features at $T = 0.05$~K, cannot be clearly resolved in this configuration.

To overcome these limitations, we employed the $E_i = 3.32$~meV configuration, which provides broader $\boldsymbol{Q}$ coverage and higher neutron flux. This enables resolution of the $\boldsymbol{Q}$ dependence of the gapped excitation centered at $\hbar\omega = 0.23(1)$~meV at $T = 0.05$~K. Figure~\ref{fig:SIFig9}a shows the powder-averaged scattering intensity as a function of energy transfer $\hbar\omega$ at $T = 0.05$~K, measured near the horizontal $(HHL)$ plane ($K \leq 0.15$~rlu along $[K\bar{K}0]$). Also shown is the intensity integrated within $\pm 0.15$~rlu around the (111) Bragg peak, scaled to match the elastic peak in Fig.~\ref{fig:SIFig9}i. The energy profile of the (111) peak provides an in-situ measure of the instrumental energy resolution, with approximately $99.49\%$ of the total intensity confined within $\hbar\omega \in [-0.2,\,0.2]$~meV and only $\sim 0.51\%$ outside this range.

In Fig.~\ref{fig:SIFig9}b, we present the $\boldsymbol{Q}$ map integrated over $\hbar\omega \in [-0.2,\,0.2]$~meV, scaled by $\sim 0.51\%$ to estimate the expected elastic leakage into the $[0.2,\,0.6]$~meV window due to finite energy resolution. Comparison with the $\boldsymbol{Q}$ map integrated over $|\hbar\omega| \in [0.2,\,0.6]$~meV [Fig.~\ref{fig:SIFig9}c] shows that the observed ``star-fish-like'' $\boldsymbol{Q}$ dependence in this energy range cannot be attributed to elastic leakage. This conclusion is supported by both the distinct momentum-space pattern—clearly different from the elastic $(100)$-type diffuse scattering—and the negligible magnitude of the estimated leakage relative to the measured signal.

For completeness, Figs.~\ref{fig:SIFig8}i,j show an analogous comparison for the $E_i = 0.99$~meV configuration between the in-plane ($K \leq 0.1$~rlu along $[K\bar{K}0]$) powder-averaged intensity and the (111) Bragg-peak profile integrated within $\pm 0.1$~rlu and scaled to match the elastic peak. This comparison further demonstrates that the gapped excitation at $\hbar\omega = 0.23(1)$~meV is well separated from the elastic line under this configuration.

\subsection{Energy-gain and energy-loss channels}\label{APP:CNCS_ConstE_Sep}
In Fig.~4d--j of the main text, we present constant-energy $\boldsymbol{Q}$ maps at $T = 0.5$~K and $T = 0.05$~K, measured with $E_i = 3.32$~meV, in which the energy-gain and energy-loss contributions are combined. We do not restrict the analysis to the energy-loss side ($\hbar\omega > 0$) with application of the correction factor $(1 + \exp(-\hbar\omega / k_{\mathrm{B}} T))$, as it is unclear whether the detailed balance relation, $I(-|\hbar\omega|, -\boldsymbol{Q}) = \exp(-\hbar\omega / k_{\mathrm{B}} T)\, I(|\hbar\omega|, \boldsymbol{Q})$, remains valid in the spin-frozen phase at $T = 0.05$~K.

To provide additional information, Fig.~\ref{fig:SIFig10} shows constant-energy $\boldsymbol{Q}$ maps measured with $E_i = 3.32$~meV, with the energy-gain and energy-loss channels displayed separately. In the positive energy-transfer range $\hbar\omega \in [0.2,\,0.6]$~meV [Fig.~\ref{fig:SIFig10}c,d], the characteristic ``star-fish-like'' magnetic scattering exhibits stronger intensity along the $[111]$ direction compared to the $[001]$ direction. In contrast, in the corresponding negative energy-transfer range $\hbar\omega \in [-0.6,\,-0.2]$~meV [Fig.~\ref{fig:SIFig10}i,j], the intensity is enhanced along the $[001]$ direction.

The origin of this asymmetry is not yet clear. It may reflect non-equilibrium spin dynamics at low temperatures in \pso, or arise from experimental factors such as imperfect background subtraction or anisotropic absorption associated with the plate-like sample geometry. Further investigation will be required to distinguish between these possibilities.

\end{document}